\documentclass[conference]{IEEEtran}
\IEEEoverridecommandlockouts
\usepackage[T1]{fontenc}
\usepackage{footnote}   
\usepackage{amsmath,amsfonts,amssymb}
\usepackage{comment}
\usepackage{lipsum}
\newtheorem{theorem}{Theorem}
\newtheorem{lemma}[theorem] {Lemma}
\usepackage{algorithm}
\usepackage{algpseudocode}
\usepackage{graphicx}
\usepackage{textcomp}
\usepackage{caption}
\usepackage{float}
\usepackage{subcaption}
\usepackage{xcolor}
\usepackage{multirow}
\usepackage{tabularx}
\usepackage{bbm,dsfont}
\newcommand{\taub}{\bar{\tau}}

\newcommand{\tauk}{{\tau}_{k}}
\newcommand{\tautil}{\tilde{\tau}}
\newcommand{\taustar}{\tau^{\ast}}

\newcommand{\Ctk}{\mathcal{C}(t_k)}
\newcommand{\Lvtau}{L_i^{v}(\tau)}
\newcommand{\Lvtauc}{L_i^{v}(\tau_c)}
\newcommand{\Lvo}{L_i^{v}(0)}

\newcommand{\Lotau}{L_0(\tau)}
\newcommand{\Lot}{L_0(0)}
\newcommand{\fng}{g(C_h)}
\newcommand{\gprime}{g'(C_h)}
\usepackage{enumerate}
\newtheorem{remark}{Remark}
\newcommand{\remove}[1]{}
\renewcommand{\baselinestretch}{0.98}
\def\BibTeX{{\rm B\kern-.05em{\sc i\kern-.025em b}\kern-.08em
    T\kern-.1667em\lower.7ex\hbox{E}\kern-.125emX}}
\begin{document}

\title{Dynamic Content caching using Restless
Multi-armed Bandits
{\footnotesize \textsuperscript{}}
\thanks{This work was supported jointly by Centre for Network Intelligence, Indian Institute of Science (IISc), a CISCO CSR initiative and Aircel TCoE project 39010C.}
}

\author{\IEEEauthorblockN{ Ankita Koley, Chandramani Singh}
\IEEEauthorblockA{\textit{Department of Electronic Systems Engineering} \\
\textit{Indian Institute of Science}\\
Bangalore 560012, India \\
Email:\{ankitakoley, chandra\}iisc.ac.in}}

\maketitle




\begin{abstract}
 We consider a dynamic content caching problem wherein the  contents get updated at a central server, and local copies of a subset of contents are cached at a local cache associated with a Base station (BS). When a content request arrives, based on whether the content is in the local cache, the BS can decide whether to fetch the content from the central server or serve the cached version from the local cache. Fetching a content incurs a fixed fetching cost, and serving the cached version incurs an ageing cost proportional to the age-of-version (AoV) of the content.
 The BS has only partial information regarding AoVs of the contents.
 We formulate an optimal content fetching and caching problem to minimize the average cost subject to cache capacity constraints. The problem suffers from the curse of dimensionality and is provably hard to solve. We formulate this problem as a continuous time restless multi-armed bandit process (RMAB), where a single content problem of the corresponding RMAB is a partially observable Markov decision process. We reformulate the single content problem as a semi-Markov decision process, prove indexability, and provide a Whittle index based solution to this problem. Finally, we compare the performance with recent work and show that our proposed policy is optimal via simulations.  
\end{abstract}



\maketitle

\section{Introduction}
Over the past few years, online social networks (OSNs) like Facebook, Instagram, LinkedIn, and YouTube have become viral platforms for users to interact, communicate, and share content over the internet. The increasing popularity has attracted many new users, resulting in a huge volume of content being shared across these platforms.  

The OSNs' Content Distribution Networks (CDNs) deploy caches in various geographical locations nearer to the users along with the central server cache to ensure the timely delivery of content; for example, the Facebook content distribution Network (FBCDN) uses several layers of caches along with backend cache~\cite{zhou2016evolution}. Along with low latency, caching reduces the backhaul traffic, reducing congestion, specifically during peak hours. Depending upon users' interaction, content relevance, location, etc., the content dynamics change, for example, Facebook's news feed or YouTube's recommendation system. As the dynamic contents get updated at the central server, these need to be replaced at the local cache of the CDNs. The server holds the most relevant version of the content for most applications and web pages~\cite{candan2001enabling}. 

The cached contents at the local cache may lose their relevance to users over time as they get updated at the central server. Upon receiving a request, whether the local cache will serve the cached content or fetch a fresh version depends upon {\it freshness} of the content. The {\it freshness} of content is measured by age-of-version (AoV), a metric proposed by  Abolhassani et al.~\cite{9488731}. AoV of content is the number of updates in the central server since that particular content is being fetched. The AoV of content depends on the age of the content and the frequency at which the content gets updated at the central server. Once a content is cached, it needs to be replaced with newer versions depending upon the AoV of the content. Otherwise, the content will become stale, resulting in an aging cost depending upon the AoV. Conversely, fetching a fresh version of the content will incur a fetching cost. Hence, designing caching policies for dynamic contents poses the following challenges:  \begin{enumerate}
\item {\it Unknown Content dynamics:} Since the contents get updated at the central server, the AoV may not be known to the local cache unless the content is fetched. 
    \item {\it Constrained cache capacity:} Local cache has smaller capacity in comparison to the central server. Hence, after fetching fresh version of the content it needs to decide whether to cache it or not based  on the the cached content and cache capacity. 
    \item {\it Dynamic requests:} Content requests vary dynamically depending on popularity, location, and many more factors. 
\end{enumerate}
Based on these factors, to minimize the cost, caching policies must carefully decide when to fetch content and whether to cache the fetched content by replacing one of the cached contents or discard the one, keeping the cached contents as they are. 
We aim to design a caching policy to minimize the average fetching and ageing costs subject to the caching constraint. This cost minimization problem falls in the restless multi-armed bandit process (RMAB) class, where we refer to each content as an arm. The single content problem of the corresponding RMAB evolves as a partially observable Markov decision process (POMDP). We further reformulate the POMDP as a semi-Markov decision process, retaining all essential information. Finally, we propose a Whittle index-based policy to solve the problem.    
\subsection{Related Work}
Kam et al.~\cite{8006505} present a framework that minimizes the cache miss rate, considering content requests affected by information freshness and popularity. However, the assumptions that there are fewer requests for packets for the content with higher age and that information freshness depends solely on the age of the content rather than the number of updates may not align well with the dynamics of today's internet. This model is further generalized by Ahanai et al.~\cite{9294151}  introducing utility functions that depend upon age and the popularity. These Age-of-information (AoI) driven models overlook the impact on freshness due to updation of contents at the central server. \\ Yates~\cite{yates2021age} introduces version AoI, a very similar metric to AoV, and studies the minimization of average version AoI in a gossiping network. Version AoI or AoV is a more suitable metric to measure freshness as it measures the number of updates since a fresh version is fetched.  However,  dealing with AoV can be challenging for most of the systems as the local cache may not be aware of the updates at the central server.    \\
Abolhassani et al.~\cite{9155324} pose two optimization frameworks. In the single-user scenario, the user checks for an update  at the central server, incurring a check cost, and then updates its cache, resulting in a cache cost. In this case, the user can store more than one content. In this model, when a request for cached content comes, the user must serve the version it currently has and the user fetches only if there is a cache miss.  
 In the multi-user scenario, each user updates its cache for free when another user requests the item via broadcast. Each user has a cache capacity of precisely one, potentially resulting in multiple copies of the same content.  
 
We consider a more flexible framework similar to~\cite{9488731}, i.e., when a request for cached content comes, the local cache can serve the version it has or request a fresh version from the server. Furthermore, in our model, the local cache does not employ a cache check; hence, it does not know when it gets updated at the server. As mentioned in~\cite{9488731}, this problem falls under the scope of a partially observable Markov decision process (POMDP) average cost problem, hard to solve. Hence, the hard cache constraint has been replaced with a probabilistic constraint to solve the problem. Further, a more flexible choice of average cache capacity constraint has been considered to satisfy the probabilistic constraint. Their solution suggests that it is enough to cache most popular items.

In a recent work, Abolhassani et al.~\cite{abolhassani2024optimal} consider a combined push and pull based caching policy, where in push based policy the central server takes the decision to update the content in the local cache, exploiting the knowledge of exact number updates of a content and in pull based caching policy the decisions are taken at the local cache exploiting the exact knowledge of request arrival. Under the assumption that once a content is stored in the local cache, it will never be discarded; push based caching policy is applied on a subset of the cached contents and pull based caching is applied for the other contents. They show that the push and pull based caching is optimal for the cached contents. However, for the uncached content, whenever there is a request there will be constant fetching cost every time. Hence, this problem needs further investigation on how to apply a push based caching when there is a provision to discard the content. However, in our work we relax the above constraint, i.e., a cached content can be discarded followed by storing a new content and  focus on pull based caching, i.e., the local cache takes the caching decisions. 

In another recent work, Abolhassani et al.~\cite{abolhassani2024swiftcache} consider a pull based caching framework with average cache constraint and study a model based and model free learning. The solution based on average cache constraint is not practical, since the actual cache capacity might be exceeded while implementing the caching policy. In comparison with~\cite{9488731} where the authors consider dynamic caching problem with a probabilistic constraint, our work focuses on developing an efficient approach to resolving the dynamic caching problem subject to a hard cache constraint.
  
\subsection{Our Contribution}
\begin{enumerate}
    \item We pose an optimal content fetching and caching problem to minimize the time average cost subject to local cache capacity constraints. This problem falls under the scope of continuous time restless multi-armed bandit process (RMAB), where the single content problem evolves as a POMDP.
    \item We reformulate the single content POMDP as a semi-Markov decision process and establish the indexability for each content. We further obtain a closed form expressions of the Whittle indices and propose the Whittle index based policy to solve the multi-content problem. 
    \item We implement the proposed Whittle index-based policy via simulations and show that the Whittle Index based policy outperforms the policy proposed by Abolhassani et al.~\cite[Theorem 2]{9488731}. We also show that the Whittle index based policy offers the same performance as the optimal policy. 
\end{enumerate}
\section{System Model} \label{sec:system model}
\begin{figure*}[t]
\centering
  \includegraphics[width=8.5 cm]{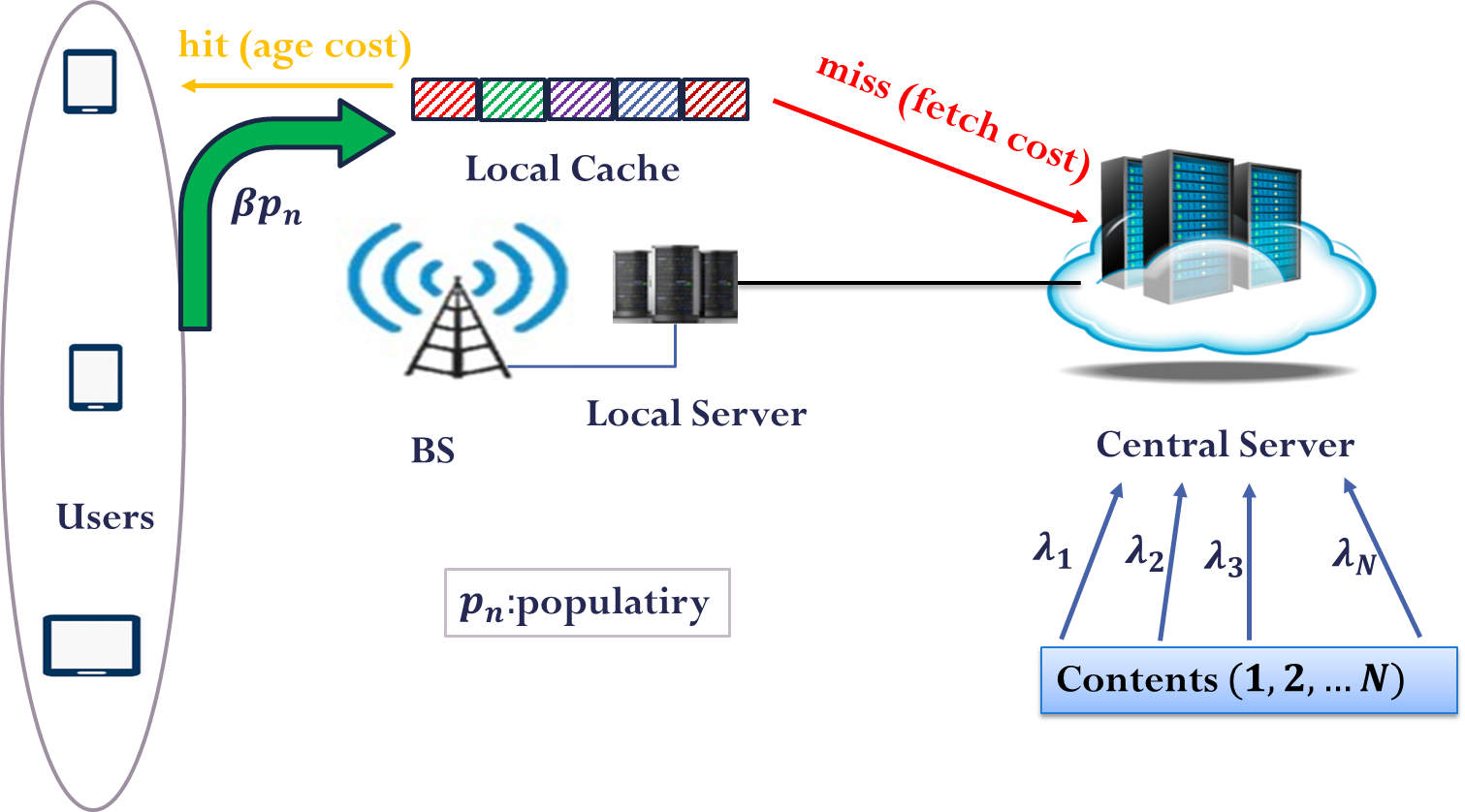}
\caption{Fresh caching of dynamic content}
\label{fig: fresh caching}
\end{figure*}

Let us consider a communication network with a central server, a Base Station (BS) associated with a local cache and an end user population. The central server hosts $N$ dynamic contents which are requested by the end users. The BS 
is connected to the central server via a wired network.
It can fetch and store up to $M$ contents  in the local cache and can locally serve these to the users upon request (Figure~\ref{fig: fresh caching}).

\paragraph*{Content dynamics} All the $N$ contents are updated according to independent Poisson processes with $\lambda_n$ being the update rate of the $n^{th}$ content. The central server always hosts the latest version of the contents. Setting $\lambda_n = 0$ for all $n$ yields the special case of {\em static} contents.    

\paragraph*{Request dynamics} The aggregate request process of the end users is a Poisson process with rate $\beta$. Each request could be for the $n^{th}$ content with probability $p_n$ independently of the other requests. Here, $p_n, 1 \leq n \leq N$ denote the relative popularity of the contents and $\sum_n p_n = 1$. For instance, the popularity of the contents on Web is widely modelled using {\em Zipf's distribution} 
wherein $p_n \propto 1/n^{\alpha} $ for the $n^{th}$ most popular content~\cite{Breslau99}. 
Under the proposed request dynamics,   $n^{th}$ content's request rate is a Poisson process with rate $p_n\beta$.
Let $\mathcal{C}(t) \subset \{1,\cdots,N\}$ denote the set of locally cached contents at time $t$. A subset of the cached contents could be updated at the central server once or multiple times since they were last fetched. Let $\nu_n(t) \geq 0$ be the number of times Content $n \in \mathcal{C}(t)$ has been updated since it was last fetched until time $t$.\footnote{We use the phrases ``$n^{th}$ content'' and ``Content $n$'' interchangeably.}  We refer to $\nu_n(t)$ as the age-of-version~(AoV) of Content $n$ at time $t$. Note that the AoVs of the cached contents are not observable at the BS. 
\paragraph*{Content fetching and ageing costs:}
If a content, say Content $n$, is requested at time $t$, one of the following scenarios may occur. 
\begin{enumerate}[(a)]
\item Content $n$ is found at the local cache. In this case, the cached copy can be served. However, the users detest receiving stale versions of the contents which is captured via {\em ageing costs}. We assume that serving a cached content incurs an ageing cost $c_a \nu_n(t)$ where $c_a$ is the ageing cost per update. Alternatively, the latest version of Content $n$ can be fetched and served, incurring a constant {\it fetching cost} $c_f$. The newly fetched copy replaces the existing one in the cache.
\item Content $n$ is not found at the local cache. It is then fetched at cost $c_f$ and served. The fetched content  either can replace an exiting content in the cache or can be discarded after serving. 
\end{enumerate}
Let $A(T)$ denote the total number of requests until time $T$. Let $t_k \geq 0$ and $e_k \in \{1,\cdots, N\}$ denote the $k^{th}$ request epoch and identity of the requested content at $t_k$, respectively. The BS does not have access to the information about the updates happening in the central server. However, the BS knows the rate $(\lambda_n)$ at which the contents are getting updated.
 Further, let $a^n_k$ denote the action at $t_k$ vis a vis Content $n \in \mathcal{C}(t_k) \cup \{e_k\}$; $a^n_k \in \{0,1\}$ if $n \in \{e_k\} \cap \mathcal{C}(t_k)$,  $a^n_k \in \{1,3\}$ if $n \in \{e_k\} \setminus \mathcal{C}(t_k)$, and $a^n_k \in \{0,2\}$ otherwise. 
\begin{table}[htbp]
\caption{Annotations of the actions}
\label{table:stateandaction}
\begin{center}
\begin{tabular}{l|l|l}
\hline
Actions & Contents & Annotations \\ 
\hline
0 & $n \in \mathcal{C}(t_k) \cap \{e_k\}$  & serve the cached  
copy and keep \\
& $n \in \mathcal{C}(t_k) \setminus \{e_k\}$ &  keep \\
\hline
1 & $n = e_k$  & fetch, serve and cache \\
\hline
2 & $n \in \mathcal{C}(t_k) \setminus \{e_k\}$ & discard \\
\hline
3 & $n = e_k$ & fetch, serve and discard \\
\hline
\end{tabular}
\label{table-action-annotation}
\end{center}
\end{table}
Hence, we define the acion set as $\mathcal{A}:=\{a\in\{0,1,2,3\}^N:\sum_{n=1}^N\mathbbm{1}_{\{a^n\in\{0,1\}\}}=M\}$
\subsection{The Optimal Content Fetching and Caching Problem} \label{sec:caching problem}
The optimal content fetching and caching problem taking actions that minimize the time average content fetching and ageing costs subject to local cache capacity constraints. It can be expressed more precisely as 
\begin{align}
\inf_{\pi\in\Pi} & \lim_{T \to \infty}\frac{1}{T}\mathbb{E}_{\pi}\left[\sum_{k=1}^{A(T)}\left(\mathbbm{1}_{\{a_k^{e_k} =  0\}} c_a \nu_{e_k}(t_k) {+} \mathbbm{1}_{\{a_k^{e_k} \in \{1,3\}\}} c_f\right)\right]\label{eq:optimal caching problem}\\
\text{s. t.} & \sum_{n=1}^N \mathbbm{1}_{\{a_{k}^n \in \{0,1\}\}} = M, \,\forall k \geq 1 \label{eq:constraint of optimal caching problem}
\end{align}
where $\Pi$ is a set of feasible policy. 
\paragraph*{Discussion} The optimal content fetching and caching problem~\eqref{eq:optimal caching problem} is a Markov decision process (MDP) with an exponential state space and action space in $M+1$. One can solve this problem using dynamic programming by restricting the states to be finite, but the solution suffers from the curse of dimensionality. However, the MDP in ~\eqref{eq:optimal caching problem} falls in the class of continuous time restless multi-armed bandit process (RMAB) and Whittle index policy~\cite{whittle1988restless} is a good heuristic solution to RMAB problems. 
\paragraph*{Whittle Index Policy}
We consider the following relaxed constraint instead of the hard constraint~\eqref{eq:constraint of optimal caching problem}:
\begin{align}
  \lim_{T\to \infty}\frac{1}{\beta T}\mathbb{E}_{\pi}\left[\sum_{k}^{A(T)}\sum_{n=1}^N \mathbbm{1}_{\{a_{k}^n \in \{0,1\}\}}\right] = M \label{eq:relaxed constraint} 
\end{align} 
Hence, we can write the Lagrangian of the problem~\eqref{eq:optimal caching problem} subject to the relaxed constraint~\eqref{eq:relaxed constraint} with multiplier $C_h$ as follows: 
\begin{align}
     \lim_{T \to \infty}&\frac{1}{T}\mathbb{E}_{\pi}\left[\sum_{k=1}^{A(T)}\left(\mathbbm{1}_{\{a_k^{e_k} =  0\}} c_a \nu_{e_k}(t_k) + \mathbbm{1}_{\{a_k^{e_k} \in \{1,3\}\}} c_f\right)\right] \nonumber\\
     &+C_h\left(\lim_{T\to \infty}\frac{1}{\beta T}\mathbb{E}_{\pi}\left[\sum_{k}^{A(T)}\sum_{n=1}^N \mathbbm{1}_{\{a_{k}^n \in \{0,1\}\}}\right] - M\right) \nonumber \\
    =\sum_{n=1}^N & \lim_{T \to \infty}\frac{1}{T}\mathbb{E}_{\pi}\left[\sum_{k}^{A(T)}\left(\mathbbm{1}_{\{a_k^{n} =  0\}} c_a \nu_{n}(t_k) {+} \mathbbm{1}_{\{a_k^{n} \in \{1,3\}\}} c_f\right)\right.\nonumber\\
    & \left. \mathbbm{1}_{\{e_k=n\}}+\frac{C_h}{\beta} \mathbbm{1}_{\{a_{k}^n \in \{0,1\}\}}\right]-C_hM \nonumber \\
      =\sum_{n=1}^N&V^n_{\pi}(C_h)-C_hM \label{eq:relaxed_problem}
\end{align} where $V^n_{\pi}(C_h)$ is the expected average cost of Content $n$ under policy $\pi$ where  a holding cost of $C_h$ per unit time is incurred for taking action $0$ or $1$, i.e., keeping or caching the Content $n$, respectively.   The above problem~\eqref{eq:relaxed_problem} can be decoupled into $n$ single content problems except the common cost part, $C_hM$ which is independent of the policy $\pi$. We define $V_n^{\pi_n(C_h)}:=\inf_{\pi \in \Pi}V_{\pi}^n(C_h)$ to be the value of the single content problem, $V_{\pi}^n(C_h)$ under the optimal policy $\pi_n(C_h)$. Let the optimal solution to the relaxed problem~\eqref{eq:relaxed_problem} be $\bar{V}=\max_{C_h}(\sum_{n=1}^N V_n^{\pi_n(C_h)}-C_hM)$. If all the contents are indexable the Whittle index policy computes the Whittle indices for each content and choose $M$ contents to cache having highest indices. We formally define indexability of a content and Whittle index policy formally in sections~\ref{sec:indexability}  and~\ref{sec:whittle index}, respectively. Indexability of a content is a property that holds if the optimal action of the single content problem $\notin \{0,1\}$, i.e., not to keep or cache the content for a given holding cost $C_h$, then  it is not  optimal to hold that content for a higher holding cost, $C'_h>C_h$. Whittle index for a content is the minimum value of $C_h$ for which it is equally attractive to keep or discard the Content $n$. We further denote $\hat{V}$ and $V_W$ to be the value of the optimal policy and the Whittle index policy of the original  problem, i.e,~\eqref{eq:optimal caching problem} under the hard constraint~\eqref{eq:constraint of optimal caching problem}. Then, it is easy to show that $\bar{V}\leq \hat{V}\leq V_W$~\cite{zhao2022multi}.  According to Whittle's conjecture~\cite{whittle1988restless} as the number of contents, $N$ grows to infinity and the buffer size, $M$ also increases proportionally with $N$, the value under Whittle index policy approaches to the value of the optimal policy of the relaxed problem, i.e., the optimal content fetching and caching problem~\eqref{eq:optimal caching problem} subject to~\eqref{eq:relaxed constraint}. \\
We note that for $C_h<0$ the optimal policy for the single content problem is to keep, i.e., $\{0,1\}$. More precisely, when the requested content is not in the cache, the optimal action is $1$ for the requested content's single content problem and the optimal action is $0$ for other contents' single content problems.  Hence, similar to~\cite{aalto2019whittle}, we will also consider $C_h\geq 0$ for our analysis. 
Going forward we will consider the single content problem for $C_h\geq 0$. We first show that the single content problem is indexable and then we compute the Whittle indices for each content.  

\begin{remark}
  Abolhassani et al.~\cite{9488731} provides an optimal solution by replacing~\eqref{eq:constraint of optimal caching problem} with a probabilistic constraint, i.e., the steady state fraction of time the items are held in the cache does not exceed the cache capacity  with high probability. To satisfy the probabilistic constraint, the authors further consider an average cache constraint and using KKT conditions they provide an optimal solution, i.e., which items to cache and for how long these should be cached. However, the constraint on average cache occupancy, $\tilde{M}=Me^{-v}$, where $v\in \mathbb{N}$ and $v$ satisfy the equation in Proposition 1 of~\cite{9488731} to ensure that the probabilistic constraint is satisfied. The  caching strategy is to cache most popular items and hold them for the times given by the solution of while the average cache occupancy is $\tilde{M}$ to ensure that that number of popular items does not exceed the actual cache capacity. The caching strategy does not cache the unpopular items at all even if the cache is not full, for each request of these unpopular items, a cost of $c_f$ incurs. This leads to under utilizing the cache capacity that further increases the average cost. However, Whittle index policy utilizes the full cache capacity and we demonstrate in section~\ref{sec:performance analysis} that it performs better than the solution provided by the authors in~~\cite{9488731}. Moreover,  we also show that the cost obtained by the Whittle index policy is very close to the optimal value of the relaxed problem, i.e., the optimal content fetching and caching problem~\eqref{eq:optimal caching problem} subject to~\eqref{eq:relaxed constraint}. This implies that the Whittle index policy offers almost the same performance as the optimal policy for the optimal content fetching and caching problem with hard cache constraints.  
\end{remark}

\section{Single Content Problem}
In this section, we discuss the single content problem with holding cost. Let $C_h \geq 0$ be the holding cost of a content per unit of time. At each epoch $t_k$ there is a request for the content with probability $p$. 
Let $A(T)$ be the number of decision epoch until time $T$. If there is a request for the  content at epoch $k$ and the content is in the cache then the following actions can be taken:
\begin{enumerate}[(a)]
    \item serve the cached copy and keep
    \item fetch serve and cache 
    \item fetch serve and discard
    \item serve the cached copy and discard. 
\end{enumerate} Recall that from the Table~\ref{table-action-annotation} the first three actions (a), (b) and (c) are $0$, $1$ and $3$, respectively. We denote the last action (d) to be $2$. We note that we did not include the action, {\it serve the cached copy and discard} in action $2$ of the multi content problem~\eqref{eq:optimal caching problem} with the constraint on the cache capacity~\eqref{eq:constraint of optimal caching problem} in section~\ref{sec:caching problem}. In the multi content problem there is no cost associated with keeping or caching a content in the cache, hence including the action, {\it serve the cached copy and discard} in action $2$ does not change the optimal solution and policy of the problem.  We aim to minimize the following average cost  for single content:\begin{align} \label{eq:single content problem}
  \lim_{T \to \infty}\frac{1}{T}\mathbb{E} &\bigg[\sum_{k=1}^{A(T)}\mathbbm{1}_{\{a_k \in \{0,2\}\}} c_a \nu(t_k) + \mathbbm{1}_{\{a_k \in \{1,3\}\}} c_f  \nonumber \\
  & +\mathbbm{1}_{\{a_k \in \{0,1\}\}}(t_{k+1}-t_k)C_h \bigg] 
\end{align}
 Instead of considering the~\eqref{eq:single content problem} we can consider the following equivalent objective~\cite{bertsekas2011dynamic} for minimization:\begin{align} \label{eq:single content problem discrete}
\inf & \lim_{T \to \infty}\frac{1}{\mathbb{E}[A(T)]}\mathbb{E}\bigg[\sum_{k=1}^{A(T)}\mathbbm{1}_{\{a_k \in \{0,2\}\}} c_a \nu(t_k) {+} \mathbbm{1}_{\{a_k \in \{1,3\}\}} c_f \nonumber\\
&+\mathbbm{1}_{\{a_k \in \{0,1\}\}}(t_{k+1}-t_k)C_h\bigg] 
\end{align} 
The above minimization problem is a partially observed Markov decision process (POMDP) where the AoV at epoch $k$, $v(t_k)$ of the content can be observed only when  the content is fetched from the central server. However, the average ageing cost at epoch $k$ can be captured via the time elapsed since it was last fetched. We denote this quantity as $\tauk:=t_k - \max\{t_l: l<k: a_{l}=1\}$. The expected AoV at epoch $k$, $\mathbb{E}[\nu(t_k)]=\lambda \tauk$. Hence, we reformulate the above POMDP problem as a semi-Markov decision process considering $\tauk$ instead of $v(t_k)$ as a part of the state. Let $b_k$ and $y_k$ be two indicator variables for the content. In each request epoch $k$, if there is a request for the content , then $b_k=1$ and $0$, otherwise. Similarly, if the content is found in the cache, then $y_k=1$ and $0$, otherwise. We denote , $s_k=\left(\tauk, y_k, b_k\right)$ if $y_k=1$ and $s_k=\left(y_k,b_k\right)$ if $y_k=0$. Hence, the state space, $\mathcal{S}:= \left\{(\tau,1,1),(\tau,1,0),(0,1),(0,0),\tau \geq 0\right\}$. The state at time $t$ are constant in between two consecutive epochs for $t_k\leq t < t_{k+1}$. The time interval between $k^{th}$ and $(k+1)^{th}$ request epoch is exponentially distributed with parameter $\beta$.\\
Given, $s_k=(\tau,1,b)$, in the $(k+1)^{th}$ epoch the states are updated as follows: 
\begin{align}
    s_{k+1}=\begin{cases}
        (\tau+\Delta \tau, 1,b'), \,\, &\text{ if } a_k=0  \\
        (\Delta \tau,1,b'), \,\, &\text{ if }   a_k=1  \\
        (0,b'), \,\, &\text{ if }a_k\in \{2,3\} \label{updation of state (tau,1,1)}
    \end{cases} 
\end{align}
where $\Delta \tau \sim \exp{(\beta)}$
and $b'\sim$ Bernoulli$(p)$. Recall from Table~\ref{table-action-annotation} that action $1$ is applicable only when the state is $(\tau,1,1)$. Given, $s_k=(0,1)$, in the $(k+1)^{th}$ epoch the states are updated as follows: 
\begin{align}
    s_{k+1}=\begin{cases}
        (\Delta \tau,1,b'), \,\, &\text{ if } a_k=1 \\
        (0,b'), \,\, &\text{ if }a_k=3
        \label{updation of state (0,1)}
    \end{cases} 
\end{align}
Recall from Table~\ref{table-action-annotation} that action $0$ and $2$ are not applicable when the state is $(0,1)$ and no action is taken when the state is $(0,0)$. 
Given,  $s_k=(0,0)$, $s_{k+1}$ will be $(0,b')$. 
The expected single stage cost of each content is given  $c(s,a)= 
   \begin{cases}
    c_a\lambda \tau\mathbbm{1}_{\{a\in \{0,2\}\}}{+}c_f\mathbbm{1}_{\{a\in\{1,3\}\}}{+}\frac{C_h}{\beta}\mathbbm{1}_{\{a\in \{0,1\}\}}, \text{if } s{=}(\tau,1,1)  \\
    c_f\mathbbm{1}_{\{a\in \{1,3\}\}}+\frac{C_h}{\beta}\mathbbm{1}_{\{a=1\}}, \text{if } s=(0,1) \\
    \frac{C_h}{\beta}\mathbbm{1}_{\{a=0\}}, \text{if } s=(\tau,1,0) \\
    0, \text{ if } s=(0,0)&
\end{cases} $ 
Hence, the cost function under an admissible policy $\pi=\{\mu_0,\mu_1,\dots\}$ and is 
\begin{align}
    J_{\pi}(s){=} & \lim_{T \to \infty}\frac{1}{\mathbb{E}_{\pi}[A(T)]}\mathbb{E}_{\pi}\left[\sum_{k=1}^{A(T)}c(s_k,a_k)\vert s_0=s\right]\label{eq:J_pi_s}\\
    J_{\ast}(s){=}&\inf_{\pi\in\Pi} \lim_{T \to \infty}{\frac{1}{\mathbb{E}_{\pi}[A(T)]}}\mathbb{E}_{\pi}\left[{\sum_{k=1}^{A(T)}}c(s_k,a_k)\vert s_0{=}s\right]\label{eq:J_star_s}
\end{align}
where $a_k=\mu_k(s_k)$ and $J_{\ast}(s)$ is the solution under optimal policy. It can be shown from ~\cite{bertsekas2011dynamic} that $J_{\pi}(s)$ and  $J_{\ast}(s)$ are independent of $s$ since the embedded discrete time Markov chains of  the problems~\eqref{eq:J_pi_s} and~\eqref{eq:J_star_s}  have  single recurrent classes. Moreover, Bellman's equation for the semi-Markov problem is similar to the discrete-time problems~\cite[Chapter~5, Section~5.3]{bertsekas2011dynamic}. 
Suppose, $h(s)$ and $\theta$ are the relative value function of state $s$ and optimal value, respectively. Let us denote $L_r(\tau)=\int_{r}^{\infty} \beta e^{-\beta t}(p h(t+\tau,1,1)  +(1-p)h(t+\tau,1,0))dt.$ Then  Bellman's equations from each state are as follows~\cite[Chapter~5, Section 5.3]{bertsekas2011dynamic}:
\begingroup
\allowdisplaybreaks
\begin{align}
     &h(\tau,1,1) =\min \left\{
     c_a\lambda \tau +\frac{C_h-\theta}{\beta}+\Lotau, c_f+\frac{C_h-\theta}{\beta} \right.\nonumber\\
     &\left.\hspace{1 cm}+\Lot, c_a\lambda \tau-\frac{\theta}{\beta}+p h(0,1)  +(1-p)h(0,0), \right. \nonumber \\
     &\left.\hspace{1 cm} c_f-\frac{\theta}{\beta}+p h(0,1) +(1-p)h(0,0)\right\} \\
   &h(\tau,1,0) {=}\min \left\{\frac{C_h-\theta}{\beta}{+}\Lotau,\right. \nonumber\\
   &\left.\hspace{1 cm}{-}\frac{\theta}{\beta}{+}p h(0,1) {+}(1-p)h(0,0) \right\} \\
  &h(0,1) {=}  \min \left\{ c_f+\frac{C_h-\theta}{\beta}+\Lot,\right.\nonumber\\ 
   &\left.\hspace{1 cm}c_f- \frac{\theta}{\beta}+ p h(0,1)+(1-p)h(0,0)\right\} \\
   &h(0,0)=-\frac{\theta}{\beta}+p h(0,1)+(1-p)h(0,0)\nonumber\\ 
   &\implies h(0,0)=-\frac{\theta}{p\beta}+h(0,1) \label{relation_h(0,0)_h(0,1)}
\end{align}
\endgroup
Using~\eqref{relation_h(0,0)_h(0,1)} we can rewrite the equations for $h(0,1)$, $h(\tau,1,0)$ and $h(\tau,1,1)$ as follows: 
\begin{align}
& h(\tau,1,1) =
     \min \left\{c_a\lambda \tau +\frac{C_h-\theta}{\beta}+\Lotau,\, \right.\nonumber\\
     &\left. c_f+\frac{C_h-\theta}{\beta}+\Lot,c_a\lambda \tau+ h(0,0), c_f+h(0,0) \right\} \,\, \label{equation_h(tau11)}\\
    & h(\tau,1,0) =\min \left\{\frac{C_h-\theta}{\beta}+\Lotau,h(0,0)\right\}   \label{equation_h(tau10)} \\
&h(0,1) =  \min \left\{c_f+\frac{C_h-\theta}{\beta}+\Lot,\, c_f+h(0,0)\right\}  \label{equation_h(0,1)}    
\end{align}
\begin{lemma}
  \begin{enumerate}[(a)] \item Both $h(\tau,1,1)$ and $h(\tau,1,0)$ are non-decreasing in $\tau$.
  \item For a given $r\geq 0$, $L_r(\tau)$ is non-decreasing in $\tau$.
  \end{enumerate} 
  \remove{
  \begin{enumerate}[(a)]
  \item $h(0,1)$ and $h(0,0)$ are constants.
    \item $h(\tau,1,1)$ and $h(\tau,1,0)$ are non-decreasing in $\tau$.
\end{enumerate} 
}
\end{lemma}
  \begin{IEEEproof}
      \begin{enumerate}[(a)]
     \item  We use induction to prove 
     the claim. The detailed proof can be found in our technical report~\cite[Appendix~A]{koley2024fresh}.
     \item Since $h(\tau,1,1)$ and $h(\tau,1,0)$ are non-decreasing in $\tau$, the integrand in $L_r(\tau)$ is also non-decreasing in $\tau$. Hence $L_r(\tau)$ is non-decreasing in $\tau$. 
     \end{enumerate}
  \end{IEEEproof}
\begin{theorem} \label{main_theorem_for_different_C_h}
The optimal policy $\pi^{\ast}$  depends on the holding cost $C_h$ and is as shown in Table~\ref{table:C_h_wise_optimal_action}, where $I=p \beta c_f-p c_a \lambda (1-e^{-\beta \tau^0}), \,\tau^0=\frac{c_f}{c_a \lambda}$, $\taustar=-\frac{1}{p\beta }+\sqrt{\left(\frac{1}{p\beta }\right)^2+\frac{2c_f}{p \beta c_a \lambda}}$ and $(\taub,\tautil)$ are the solutions to the following two equations:
\begin{align}
    c_a \lambda p \beta (\tautil \taub-\frac{\taub^2}{2})-C_h \taub+c_a \lambda \tautil -c_f&=0 \label{eq:1_for_uniqueness}\\
     \beta (\tautil-\taub)+e^{-\beta (\tautil-\taub)}-1-\frac{C_h}{ p c_a \lambda}&=0 \label{eq:2_for_uniqueness}
\end{align}
Additionally, $0\leq \taub < \taustar <\tautil \leq \tau^0.$
\begin{table}[htbp]
\centering
\caption{State-wise optimal actions for different values of $C_h$ }
\setlength\tabcolsep{3pt}
\label{table:C_h_wise_optimal_action}
\begin{tabular}{|c|c|c|c|c|}
\hline
   Holding   &  \multicolumn{3}{|c|}{Optimal policy ($\pi^\ast$)} & Optimal \\ 
     \cline{2-4}
    {cost ($C_h$)} & $\pi^{\ast}\left(\tau,1,1\right)$ & $\pi^{\ast}\left(\tau,1,0\right)$ & $\pi^{\ast}\left( 0,1\right)$&cost $(\theta)$ \\
     \hline 
      $C_h = 0$ &$
         0\text{ for }\tau \leq \taustar 
           $ & $0$  &  $1$ &$p\beta c_a \lambda \taustar$ \\
           &$1\text{ for } \tau > \taustar$ & & & \\ 
      \hline
       $0< C_h < I$& $
          0\text{ for } \tau \leq \taub
          $ & $
          0\text{  for } \tau \leq \taub  
      $  &  $1$ & $p\beta c_a \lambda \tautil$ \\
        & $2$\text{ for } $\taub \leq \tau \leq \tautil$&  $ 2\text{ for }  \tau >\taub$& & \\ 
        & $1$\text{ for } $\tau > \tautil$ & & & \\
      \hline
      $C_h>I$ & $
          2\text{ for } \tau \leq \tau^0 
      $ & $2$ & $3$ &$p\beta c_f$\\
      &$3$\text{ for } $\tau > \tau^0$ & & & \\ 
      \hline
\end{tabular}
\end{table}
\end{theorem}
\begin{remark}
Theorem~\ref{main_theorem_for_different_C_h} establishes that the optimal policy for the single user problem is of threshold type for a given value of $C_h$ where the threshold is on the time elapsed since a fresh version of the content is fetched ($\tau$). Moreover, it provides the closed forms of thresholds for different values of $C_h$ in Table~\ref{table:C_h_wise_optimal_action}. 
\end{remark}
\begin{IEEEproof}
   We outline the proof of this theorem here, the detailed proof can be found in our technical report~\cite[Appendix~B]{koley2024fresh}.
   \begin{enumerate} [(a)]
       \item We consider three  disjoint cases conditioned on the values of $h(\tau,1,1)$ and $h(\tau,1,0)$ as follows:
       \begin{enumerate}[I.]
           \item   $\frac{C_h-\theta}{\beta}+\Lot\geq h(0,0)$.          
           \item  {Since $h(\tau,1,1)$ and $h(\tau,1,0)$ are non-decreasing in $\tau$, the complement of case~$1$ is      there exists a $\taub > 0$ such that $\taub= \min \left\{\tau > 0:\frac{C_h-\theta}{\beta}+\Lotau
    \geq h(0,0)\right \}$. Since $\Lotau$ is non-decreasing in $\tau$, $c_a\lambda\tau+\Lotau$ is strictly increasing in $\tau$. Since, $c_f+\Lot$ is a constant and $c_a\lambda \tau+\Lotau <c_f+\Lot$ at $\tau=0$, there exists a $\taustar>0$ such that $c_a\lambda \taustar+L_0(\taustar)=c_f+\Lot$.
     Hence, $\taustar= \min \left\{\tau \geq 0 :c_a\lambda \tau + \Lotau = c_f+\Lot\right\} $. There could be two cases $\taustar \leq \taub$ or $\taustar\geq\taub$. We consider $\taustar\leq\taub$ in this case and $\taustar>\taub$ in the subsequent case. }
    \item  There exists a $\taub > 0$ such that $\taub= \min \left\{\tau>0:\frac{C_h-\theta}{\beta}+\Lotau
    \geq h(0,0)\right \} $ and $\taustar>\taub$.  In this case, for $\tau \leq \taub$, $C_a \lambda \tau+\frac{C_h-\theta}{\beta}+\Lotau \leq C_a \lambda \tau+h(0,0)$ and $\tau=\taub$, \begin{align}
      C_a \lambda \tau+\frac{C_h-\theta}{\beta}+\Lotau = C_a \lambda \tau+h(0,0) \label{equation_for slope_of_L.H.S and R.H.S}  
      \end{align}
      Since $L_0(\tau)$ is non-decreasing in $\tau$, both R.H.S and L.H.S in~\eqref{equation_for slope_of_L.H.S and R.H.S} are increasing in $\tau$. 
    The slope of R.H.S is $C_a \lambda$ and the slope of L.H.S is at least $C_a \lambda$, i.e., the slope of L.H.S is greater than or equal to that of R.H.S.  Since at $\taustar$, the  L.H.S exceeds $c_f+\frac{C_h-\theta}{\beta}+\Lot$, it is obvious that there exists a $\tautil \geq \taustar$ such that, R.H.S exceeds $c_f+\frac{C_h-\theta}{\beta}+\Lot$. This implies  that there exists  a $\tilde{\tau} > \bar{\tau}$ such that $\tautil=\min\left\{\tau > \taub :c_a\lambda \tau + h(0,0) = c_f+\frac{C_h-\theta}{\beta}+\Lot\right\} $.\\ 
    Now, we will show that suppose $\exists$ a $\tautil$ as above, then $\taustar>\taub$. Since at $\tautil$ R.H.S exceeds $c_f+\frac{C_h-\theta}{\beta}+\Lot$  and slope of R.H.S is no more than the L.H.S, the L.H.S exceeds  $c_f+\frac{C_h-\theta}{\beta}+\Lot$ at a value of $\tau$ no more than $\taub$. Hence, $\taustar> \taub$. 
    Hence, we consider the last case as follows:  there exists $\taub > 0$ such that $\taub= \min \left\{\tau>0:\frac{C_h-\theta}{\beta}+\Lotau
    \geq h(0,0)\right \} $ and there exists  a $\tilde{\tau} > \bar{\tau}$ such that $\tautil=\min\left\{\tau > \taub:c_a\lambda \tau + h(0,0) = c_f+\frac{C_h-\theta}{\beta}+\Lot \right\}$.
       \end{enumerate}
  
       \item We derive the optimal actions in each of the cases. Furthermore, we  show that Case~I holds iff $C_h>I$, Case~II holds iff $C_h=0$, and  Case~III holds iff $0<C_h\leq I$. This also implies that the above mentioned three cases are mutually exclusive and exhaustive. 
       \item Recall $\tau^{0}=\frac{c_f}{c_a\lambda}$ theorem~\ref{main_theorem_for_different_C_h}. Hence,  from case~III. $c_a\lambda\tau^{0}+h(0,0)=c_f+h(0,0)\geq c_f+\frac{C_h-\theta}{\beta}+\Lot$. Hence, by the definition of $\tautil$, $\tautil \leq \tau^{0}$. Hence, from the Lemma~\ref{lemma:uniqueness} $\taub\leq\taustar\leq\tautil\leq\tau^{0}$. 
   \end{enumerate} 
           \end{IEEEproof}
 \begin{lemma} 
 \label{lemma:uniqueness}
 \begin{enumerate}
     \item \eqref{eq:2_for_uniqueness} and~\eqref{eq:1_for_uniqueness} are satisfied for unique non-negative values of $\tautil$ and $\taub$. 
     \item  $\taub$ and $\tautil$ are decreasing and increasing function of $C_h$, for $0< C_h\leq I$ respectively. Furthermore, $\taub\leq\taustar\leq\tautil$. 
 \end{enumerate}
\end{lemma}
\begin{IEEEproof}
    See {\it Proof of Lemma~\ref{lemma:uniqueness}} in our technical report~\cite[Appendix~E]{koley2024fresh}. 
     
\end{IEEEproof}

\begin{figure}
\begin{subfigure}[t]{0.5\linewidth}
\centering  
\includegraphics[width=\linewidth]{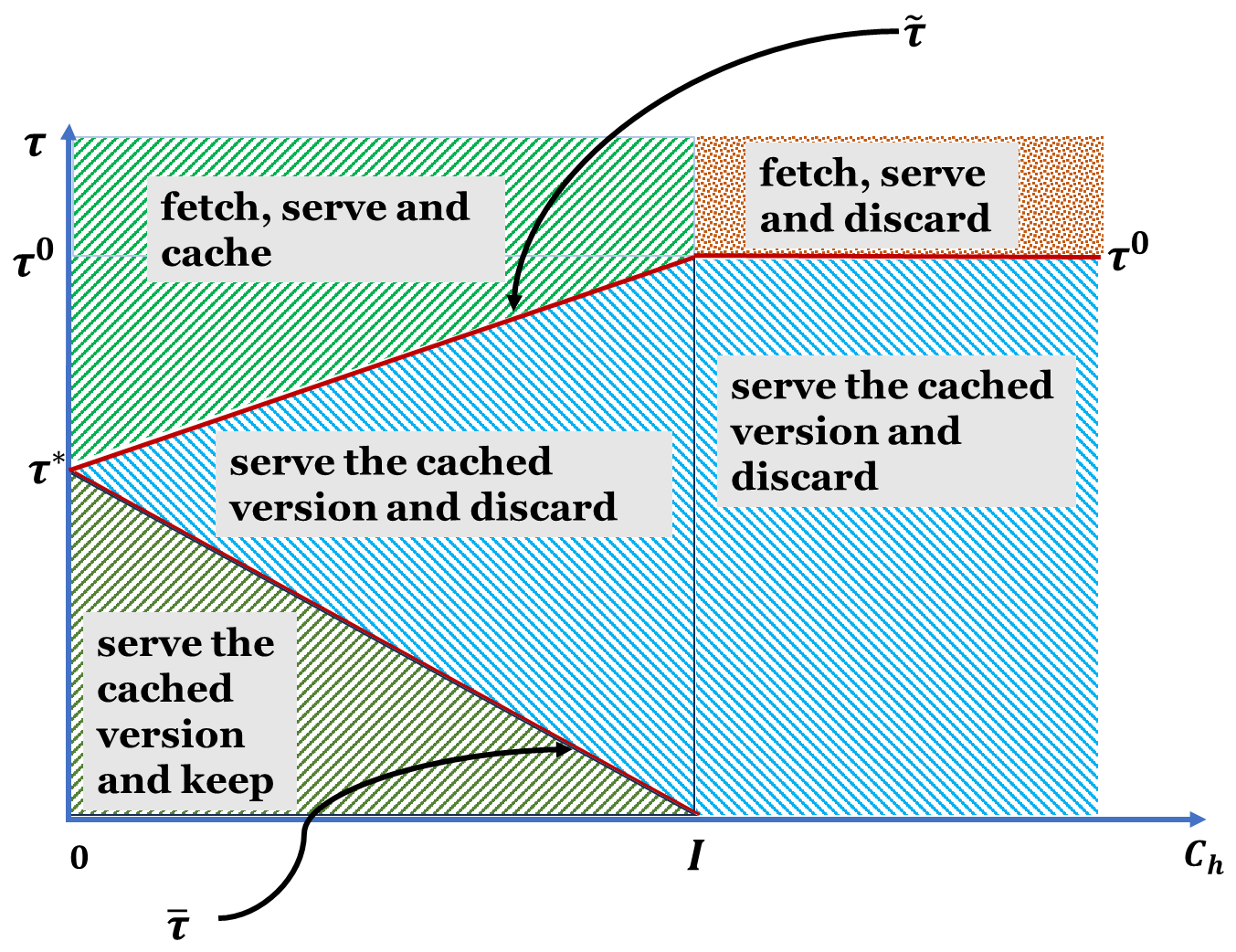}
   \caption{$s=(\tau,1,1)$}
   \label{fig:tau_vs_C_h_one_one}
\end{subfigure}
\begin{subfigure}[t]{0.49\linewidth}
\centering  
\includegraphics[width=\linewidth]{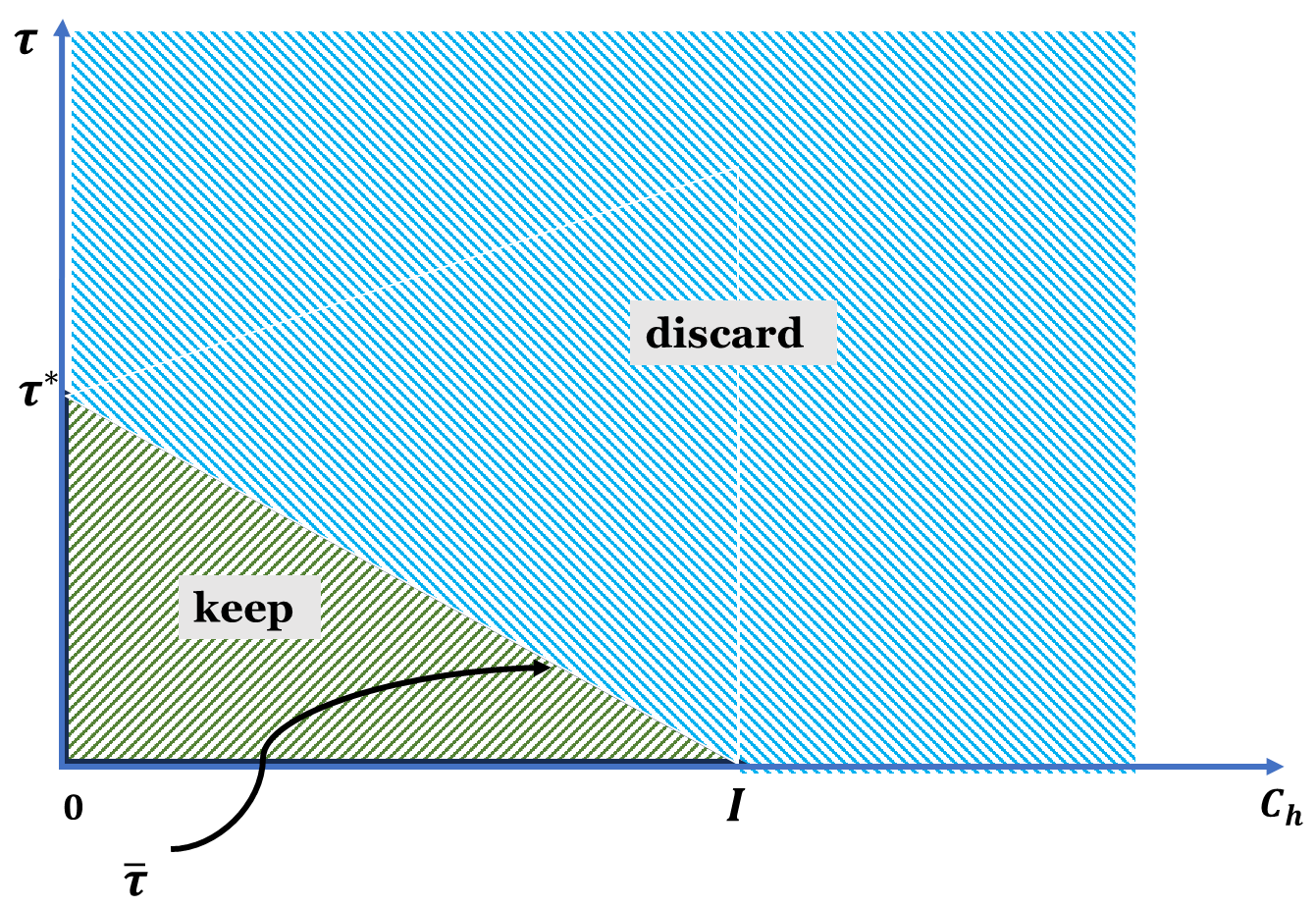} 
   \caption{$s=(\tau,1,0)$}
   \label{fig:tau_vs_C_h_one_zero}
\end{subfigure}
\caption{Optimal policy structure with respect to $C_h$ for the state $(\tau,1,1)$ and $(\tau,1,0)$ where $\taustar,\tau^0$ and  $I$ are as in Theorem~\ref{main_theorem_for_different_C_h}.}
\label{fig: tau vs C_h}
\end{figure}
\begin{remark}
  In Figures~\ref{fig:tau_vs_C_h_one_one} and~\ref{fig:tau_vs_C_h_one_zero}, we plot the threshold values of $\tau$ as we vary $C_h$ and indicate various regions depending upon optimal actions for the states $\{(\tau,1,1):\tau \geq 0\}$ and $\{(\tau,1,0):\tau \geq 0\}$, respectively. We describe the actions for different states as follows:\begin{enumerate}[(a)]
       \item When the state of the content is $(\tau,1,1)$, given $C_h=0$, the optimal action is to serve the cached version and keep for $\tau\leq\taustar$ and fetch, serve, and keep for $\tau>\taustar$. For $0<C_h <I$, the optimal action is to serve the cached version and keep for $\tau\leq\taub$,  serve the cached version and discard for $\taub <\tau\leq\tautil$, and fetch, serve and cache for $\tau>\tautil$. For $C_h\geq I$, the optimal action is to serve the cached version and discard for $\tau\leq \tau^{0}$, and fetch, serve, and cache for $\tau>\tau^{0}$. 
        We observe in Figure~\ref{fig:tau_vs_C_h_one_one} that the area region where the optimal action to serve the cached version and discard increases as $C_h$ increases and becomes constant after $C_h$ exceeds $I$. Furthermore, the region where the optimal action is to discard, i.e., $ \{2,3\}$, expands as $C_h$ increases from $0$ to $I$, and the region extends to infinity after $C_h$ exceeds $I$.   
       \item When the state of the content is $(\tau,1,0)$, given $C_h=0$, keep for regardless of the value of $\tau$. For $0<C_h <I$, the optimal action is to  keep for $\tau\leq\taub$ and discard for $\tau>\taub$. For  $C_h \geq I$, the optimal action is to discard for any value of $\tau$. We observe in Figure~\ref{fig:tau_vs_C_h_one_zero} as we increase $C_h$, the area of the region where the optimal action is to discard expands and extends t infinity as $C_h$ exceeds $I$.  
      \item When the state of the content is $(0,1)$,  the optimal action is to fetch and fetch serve and discard for $C_h<I$ and $C_h\geq I$, respectively.
   \end{enumerate} 
\end{remark}
In the next section, we establish 
{\em indexability} of the single content problem.

 \subsection{Indexability}\label{sec:indexability}

In this section first we show that the single content MDP is indexable and then we compute the Whittle index for each content $n \in \mathcal{C}(t_k)\cup e_k$. 
The passive set under the holding cost $C_h$ is given by the states for which it is optimal to discard the content and more precisely:
$\mathcal{P}(C_h):=\left\{s: \pi^{\ast}(s) \in \{2,3\}\right\}$. \\
{\it Definition of Indexability:} A content is indexable if  the passive set of that content satisfies the following conditions~\cite{8919842}:
\begin{enumerate}[(a)]
    \item $\mathcal{P}(0)=\emptyset$ and $\mathcal{P}(\infty)=\mathcal{S}$
    \item $\mathcal{P}(C_{h_1}) \subseteq \mathcal{P}(C_{h_2})$ for $C_{h_1} \leq C_{h_2}$
\end{enumerate}
 The corresponding RMAB is indexable if the single content MDP is indexable.
 \begin{theorem} \label{thm:indexability}
     The single content MDP is indexable.
 \end{theorem}
 \begin{IEEEproof}
Recall from Table~\ref{table:C_h_wise_optimal_action} if $C_h=0$, the optimal action $\pi^{\ast}(s) \in \{0,1\}$ $\forall \, s$, i.e., it is optimal to keep the content for any state. Hence,     $\mathcal{P}(0)=\emptyset$. \\
If $C_h\geq I$, then the optimal action $\pi^{\ast}(s) \in \{2,3\}$ $\forall \,s$, i.e., it is optimal discard the content for any state. Hence, $\mathcal{P}(C_h)=\mathcal{S}$ for and $C_h\geq I$ and moreover $\mathcal{P}(\infty)=\mathcal{S}$. Hence, for $C_h\geq I$, $\mathcal{P}(C_{h_1})= \mathcal{P}(C_{h_2})$. It is remaining to show that $\mathcal{P}(C_{h_1})\subseteq \mathcal{P}(C_{h_2})$ for $ 0 < C_{h_1}\leq C_{h_2} < I$. Recall, from Lemma~\ref{lemma:uniqueness} that  $(\taub_{C_h},\tautil_{C_h})$ are the unique solutions to the following two equations:
\begin{itemize}
    \item[1.] $c_a \lambda p \beta (\tautil_{C_h} \taub_{C_h}-\frac{\taub_{C_h}^2}{2})-C_h \taub_{C_h}+c_a \lambda \tautil_{C_h} -c_f=0$
    \item[2.] $\beta (\tautil_{C_h}-\taub_{C_h})+e^{-\beta (\tautil_{C_h}-\taub_{C_h})}-1-\frac{C_h}{ p c_a \lambda}=0$
\end{itemize}  
Furthermore,  from Lemma~\ref{lemma:uniqueness} $\tautil$ and $\taub$ increases and decreases, respectively as we increase $C_h$ . 
Hence, for $0 < C_{h_1} \leq C_{h_2} < I$, we note that $\taub_{C_{h_2}}\leq \taub_{C_{h_1}}<\tautil_{C_{h_1}}\leq \tautil_{C_{h_2}}$. The passive sets corresponding to $C_{h_1}$ and $C_{h_1}$ are $\mathcal{P}(C_{h_1})=\{(\tau,1,1): \taub_{C_{h_1}}\leq \tau \leq \tautil_{C_{h_1}}\}\cup \{(\tau',1,0):\tau'\geq \taub_{C_{h_1}} \}\cup \{(0,0)\}$ and $\mathcal{P}(C_{h_2})=\{(\tau,1,1): \taub_{C_{h_2}}\leq \tau \leq \tautil_{C_{h_2}}\}\cup \{(\tau',1,0):\tau'\geq \taub_{C_{h_2}} \}\cup \{(0,0)\}$,  respectively. Hence, $\mathcal{P}(C_{h_1})\subseteq \mathcal{P}(C_{h_2})$ for $C_{h_1}\leq C_{h_2}.$
 \end{IEEEproof}
 In the following we consider the Multi-content problem and obtain the Whittle indices for each content. Finally, we propose a Whittle index based policy as a solution to the problem~\eqref{eq:optimal caching problem} subject to~\eqref{eq:constraint of optimal caching problem}. 
\subsection{ Whittle Index policy for Multi-content problem}
\label{sec:whittle index}
  If a content is indexable, its Whittle index $W(s)$
associated with state $s$ is the minimum cost that moves this state from the
active set to the passive set. Equivalently,
$W(s)=\min\{C_h:s \in \mathcal{P}(C_h)\}$.  \\
We already mentioned in Section~\ref{sec:indexability} that the single content MDP is indexable and hence the RMAB is also indexable. We obtained the closed form solution of the Whittle index for the contents $n \in \Ctk \cup \{e_k\}$. The Content $n \in \Ctk \cup \{e_k\}$ can be in the following states as follows: 
\begin{align}
    s_n(t_k)=\begin{cases}
        (\tauk^{n},1,1), \text{ if } n \in \Ctk \cap \{e_k\} \\
        (\tau_k^{n},1,0) \text{ if } n \in \Ctk \setminus \{e_k\} \\
        (0,1) \text{ if } n \in \{e_k\} \setminus \Ctk  
    \end{cases}
    \label{eq:states_multicontent}
\end{align}
The Whittle index policy computes the indices for the contents in $C(t_k)\cup e_k$ and keep $M$ contents in the cache having highest indices.   
As discussed in Section~\ref{sec:system model} one of the following scenarios may occur. 
\begin{enumerate}[(a)]
\item \label{policy_part_a}Suppose, the requested content is in the cache, i.e., $e_k\in\mathcal{C}(t_k)$. We note that $\vert e_k\cup \mathcal{C}(t_k) \vert=M$ and cache size is $M$, no content needs to discarded.  Hence, there are no need to to compute the Whittle indices for each content.  In this case the optimal action $\pi^{\ast}(s_k)\in \{0,1\}$ for  Content $e_k$ and for other contents $\mathcal{C}(t_k)\setminus e_k$, the optimal action is $0$. Finally, from Table~\ref{table:C_h_wise_optimal_action}, we consider the following actions for Content $e_k$:
 \begin{enumerate}[(a)]
    \item 0 or serve the cached copy if $\tau \leq \tau^{e_k \ast}$
    \item 1 or fetch and cache if $\tau > \tau^{e_k \ast}$
\end{enumerate}
where  $\tau^{e_k \ast}=-\frac{1}{p_{e_k}\beta }+\sqrt{(\frac{1}{p_{e_k}\beta })^2+\frac{2c_f}{p_{e_k} \beta c_a \lambda_{e_k}}}$. Since, we do not discard any content in this case and Content $e_k$ is in state $(\tauk^{e_k},1,1)$, we consider the first case in Table~\ref{table:C_h_wise_optimal_action}, where the optimal action is not to discard for state $(\tauk^{e_k},1,1)$. 
\item Let us consider the case when the requested content is not in the cache, i.e., $e_k\notin \mathcal{C}(t_k)$. It is then fetched at cost $c_f$ and served. We observe that, $\vert \mathcal{C}(t_k) \cup e_k \vert = M+1$. Among $M+1$ contents, one content need to be discarded. The fetched content either can replace an existing content in the cache or can be discarded after serving. We calculate the Whittle indices for each content in $ \mathcal{C}(t_k) \cup e_k$ and discard the content having least Whittle index. Content $n$ is in state $(\tauk^{n},1,0)$ if $n\in\mathcal{C}(t_k)\setminus e_k$ and $(0,1)$ if $n=e_k$. Let us denote $W_n(s)$ as Whittle index for content $n$ at state $s$. Then,
\begin{theorem} \label{thm:whittle indices}
For $n\in \mathcal{C}(t_k)\setminus e_k$,
\begin{align}
    W_n((\tau,1,0))= \begin{cases}
    C_h(\tau) &\text{ if } \tau < \tau^{n\ast}\\
        0 , &\text{ if }\tau \geq \tau^{n \ast}       
    \end{cases}
\end{align}
Where $\left(C_h(\tau), \tautil(\tau)\right)$ be the unique solution to the following to equations:
\begin{align}
c_a \lambda_n p_n \beta (\tautil \tau-\frac{\tau^2}{2})-C_h \tau+c_a \lambda_n \tautil -c_f&=0 \label{first_equation_of_whittle_index}\\
    \beta (\tautil-\tau)+e^{-\beta (\tautil-\tau)}-1-\frac{ C_h}{ p_n c_a \lambda_n}&=0  \label{second_equation_of_whittle_index}
\end{align}
and $W_{e_k}((0,1))=I_{e_k}=p_{e_k} \beta c_f-p_{e_k} c_a \lambda_{e_k} (1-e^{- \frac{\beta c_f}{c_a\lambda_{e_k}}})$. 
\end{theorem}
\begin{IEEEproof}
  Suppose, the state of Content $n$ is $(\tau,1,0)$. Then the Whittle index will be minimum $C_h$ for which $\pi^{\ast}(\tau,1,0)=2.$  From Table ~\ref{table:C_h_wise_optimal_action}, we note that the optimal action is $2$ for some $C_h\in (0,I_n]$, where $I_n=p_{n} \beta c_f-p_{n} c_a \lambda_{n} (1-e^{- \frac{\beta c_f}{c_a\lambda_{n}}})$. From Theorem~\ref{main_theorem_for_different_C_h}, the Whittle index will be $\min\left\{C_h: C_h \text{ satisfies}~\eqref{first_equation_of_whittle_index} \text{ and}~\eqref{second_equation_of_whittle_index}\right\}$. We note that~\eqref{first_equation_of_whittle_index} and~\eqref{second_equation_of_whittle_index} are same as~\eqref{eq:1_for_uniqueness} and~\eqref{eq:2_for_uniqueness}, respectively except $p$ being replaced by $p_n$ and $\lambda$ being replaced by $\lambda_n$.   From Lemma~\ref{lemma:uniqueness}, we note that there is an unique value of $C_h$ for which~\eqref{first_equation_of_whittle_index} and~\eqref{second_equation_of_whittle_index} are satisfied. Hence, we obtain the Whittle indices for contents in $ \mathcal{C}(t_k)\setminus e_k$. The content $e_k$ has state $(0,1)$. We note that from Table~\ref{table:C_h_wise_optimal_action}, for $C_h<I_{e_k}$, the optimal action is $1$ or fetch, serve and cache and for $C_h\geq I_{e_k}$, the optimal action is $3$ or fetch, serve and discard. Hence the Whittle index for $e_k$ is $I_{e_k}$.
\end{IEEEproof}

We propose the Algorithm~$1$ based on Whittle Index as a solution to the problem~\eqref{eq:optimal caching problem}.  
\end{enumerate}
\begin{algorithm}
\caption{Whittle Index-based Caching for Freshness}
\label{alg:equivalentmap}
\begin{algorithmic}[1]
\State Initialization: Start with a random cache and random vector $\{\tauk^{n}:n\in N\}$
\Procedure{Caching}{ }
    \ForAll{request epoch}
              \If {the requested content is found in the cache, i.e., $e_k \in \mathcal{C}(t_k)$} 
                     \State do not discard any content
                      \If{$\tauk^{e_k}$ exceeds the threshold, $\tauk^{e_k*}$}  
                          \State   fetch and cache the requested content 
                       \Else
                           \State  serve and keep
                      \EndIf

             \Else
                    \State fetch the requested content
                     \ForAll{ contents in the cache and requested content}
                         \State calculate Whittle indices
                      \EndFor
                  \State  Discard the content having the least Whittle index
             \EndIf
   \EndFor
\EndProcedure
\end{algorithmic}
\end{algorithm}

\subsection{Whittle Index Policy for Multi-content having Different Sizes}
In this section we consider a more general problem where the contents can be of different sizes. Let $d_n$ be the size of content $n$. Then the fetching cost of item $n$ is $c_fd_n$, where $c_f$ is the fetching cost of an item of unit size. We reformulate the optimal caching problem~\eqref{eq:optimal caching problem} as follows:
\begin{align}
\inf_{\pi\in\Pi} & \lim_{T \to \infty}\frac{1}{T}{\mathbb{E}_{\pi}}\left[{\sum_{k=1}^{A(T)}}\mathbbm{1}_{\{a_k^{e_k} =  0\}} c_a \nu_{e_k}{(t_k)} {+} \mathbbm{1}_{\{a_k^{e_k} \in \{1,3\}\}} c_fd_{e_k}\right] \nonumber\\
\text{s. t.} & \sum_{n=1}^N d_n\mathbbm{1}_{\{a_{k}^n \in \{0,1\}\}} \leq M, \,\forall k \geq 1 \label{eq:diff_size_constraint of optimal caching problem}
\end{align}
Using similar computations as~\eqref{eq:relaxed constraint},~\eqref{eq:single content problem}and~\eqref{eq:single content problem discrete} we can write the single content problem as: 
\begin{align}
\inf & \lim_{T \to \infty}\frac{1}{\mathbb{E}[A(T)]}\mathbb{E}\bigg[\sum_{k=1}^{A(T)}\mathbbm{1}_{\{a_k \in \{0,2\}\}} c_a \nu(t_k) {+} \mathbbm{1}_{\{a_k \in \{1,3\}\}} dc_f \nonumber\\
&+\mathbbm{1}_{\{a_k \in \{0,1\}\}}(t_{k+1}-t_k)dC_h\bigg] 
\end{align}
where $d$ replaces $d_n$ in~\eqref{eq:diff_size_constraint of optimal caching problem}. For the single content problem also we can obtain similar result as Theorem~\ref{main_theorem_for_different_C_h} where $C_h$ is replaced by $dC_h$ and $c_f$ is replaced by $dc_f$ and prove that the single content MDP is indexable. For the multi-content problem the policy would be as follows:
\begin{enumerate}[(a)]
    \item Suppose, the requested content is found in the cache, i.e., $e_k\in \Ctk$. Then we consider the following actions for Content $e_k$:
 \begin{enumerate}[(a)]
    \item 0 or serve the cached copy if $\tau \leq \tau_1^{e_k \ast}$
    \item 1 or fetch and cache if $\tau > \tau_1^{e_k \ast}$
\end{enumerate}
where  $\tau_1^{e_k \ast}=-\frac{1}{p_{e_k}\beta }+\sqrt{(\frac{1}{p_{e_k}\beta })^2+\frac{2c_fd_{e_k}}{p_{e_k} \beta c_a \lambda_{e_k}}}$. We observe that $\tau_1^{e_k \ast}$ has the same expression as $\tau^{e_k\ast}$ in~Algorithm $1$ except $c_f$ is replaced by $c_fd_{e_k}.$
\item Suppose, the requested content is not found in the cache, i.e., $e_k\notin \Ctk$. Then, we compute the Whittle indices of $M+1$ contents and choose a number of top contents until the constraint~\eqref{eq:diff_size_constraint of optimal caching problem} is violated~\cite{7593334}.  
\end{enumerate}
 \section{Performance Analysis}\label{sec:performance analysis}
In this section, we study the performance of the Whittle Index based policy. We consider the number of contents, $N=1000$, and study the impact on average cost as we increase the cache size from $40$ to $100$ under the following settings \cite{9488731},~\cite{9771060}: The popularity of the content $n$ be $p_n=\frac{1}{n}$, the fetching cost and the aging cost are $c_f=1$ and $c_a=0.1$, respectively, and the arrival rate, $\beta=5$.
\subsection{Comparison}
\begin{figure}[ht]
  \centering
\includegraphics[scale=.22]{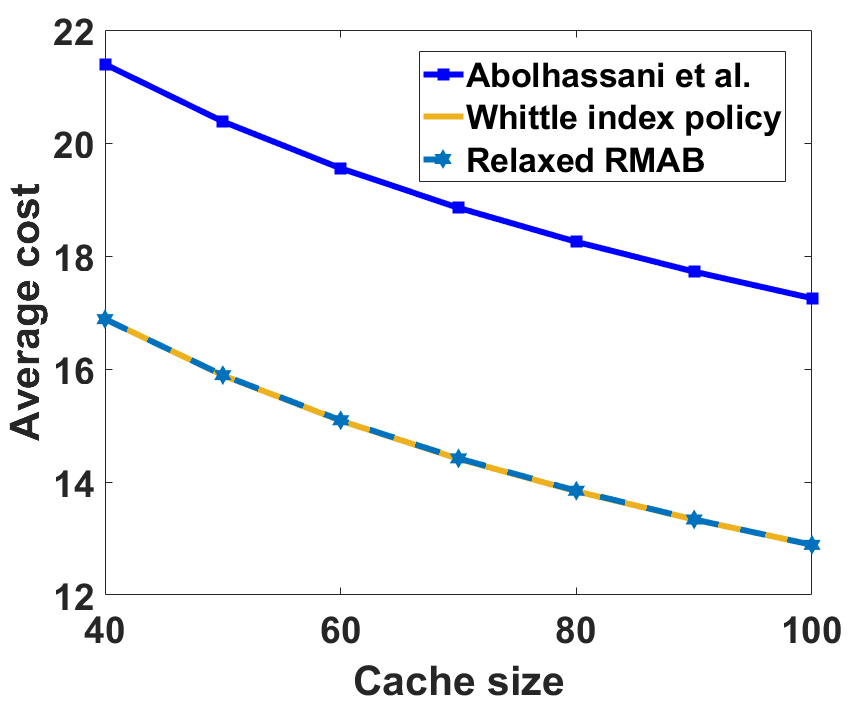}
\caption{Comparison of average cost between Whittle Index based policy and the policy by Abolhassani et al.\cite{9488731}}
\label{fig: comparison with infocom paper}
\end{figure}

The update rate of the content $n$ at the central server is $\lambda_n=\lambda=0.01$. We implement the the proposed Whittle index based policy and plot the average cost for different cache sizes. Figure~\ref{fig: comparison with infocom paper} shows that the Whittle Index based policy outperforms the policy proposed by Abolhassani et al.~\cite[Theorem~2]{9488731}. We numerically compute the average cost of the relaxed RMAB problem using Table~\ref{table:C_h_wise_optimal_action}. Furthermore, the average cost under the Whittle index policy is almost same as the optimal cost of the relaxed RMAB problem. As we mentioned earlier the optimal cost of the relaxed RMAB problem acts as a lower bound of the optimal cost of the original problem, the Whittle index policy is therefore  optimal.    
\subsection{Effect of update rate ($\lambda$) on average cost}
\begin{figure}
\centering
\begin{subfigure}{0.49\linewidth}
{\includegraphics[width=1\linewidth]{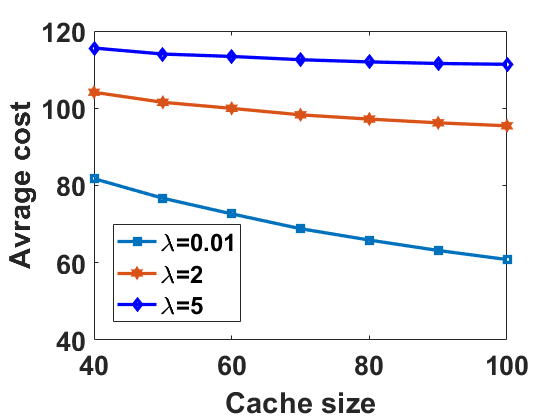}}
\caption{Effect of update rate ($\lambda$) on the average cost }
  \label{fig: comparison with varying lambda}
  \end{subfigure}
  \begin{subfigure}{0.49\linewidth}
{\includegraphics[width=1\linewidth]{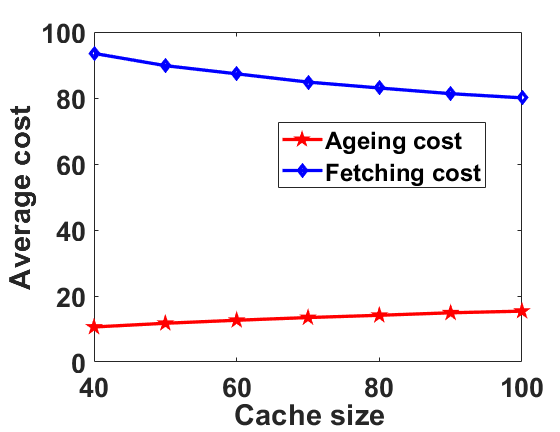}}
\caption{Effect of Cache size on average ageing cost and fetching cost}
  \label{fig: comparison between age and fetch cost}
  \end{subfigure}
\caption{Effect of update rate and cache size on average ageing cost and fetching cost }
\end{figure}

In this subsection, we implement the Whittle index policy to  compute the average cost for three different update rates $\lambda=0.01, 2$ and $5$ and for $c_f=5$, while keeping all the other parameters the same as in section~\ref{sec:performance analysis}. We observe that the average cost increases as the update rate increases (see Figure~\ref{fig: comparison with varying lambda}). Since the ageing cost of a content is $\propto$ its update rate, it is obvious the average cost will increase as we increase the update rate. 
Moreover, the average cost decreases as the cache size increases (see Figure~\ref{fig: comparison with varying lambda}). This is because a larger cache size reduces fetching costs, given that more items can be accommodated. However, lowering fetches results in serving more items with older versions, therefore causing an increase in the aging cost. For example, see Figure~\ref{fig: comparison between age and fetch cost}, where we plot the average ageing cost and fetching cost for $\lambda=2$. The overall average cost decreases due to the Whittle index based policy's management of the trade-off between fetching costs and aging costs. 

\subsection{Effect of fetching cost ($c_f$) on average cost and number of fetches during cache hits}
\begin{figure}[ht]
  \centering
\includegraphics[scale=.35]{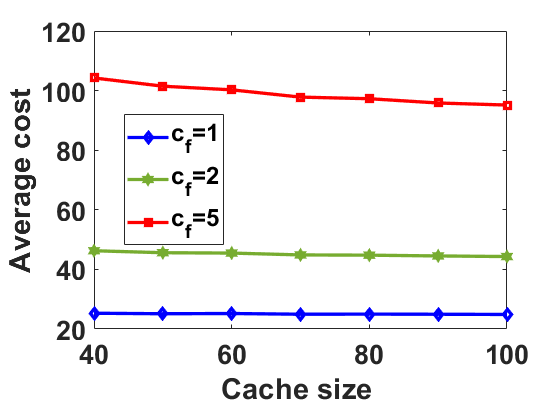}
\caption{Effect of fetching cost $(c_f)$ on average cost }
\label{fig: comparison between hit percentage}
\end{figure}

In this subsection we will discuss how the fetching cost affects number of fetches during cache hits. We implement the Whittle index based policy  for different fetching costs, $c_f=1,2,5$ and for $\lambda=2$, while keeping all the other parameters the same as in section~\ref{sec:performance analysis}. 
Since the average fetching cost increases as $c_f$ increases, we observe that ( see Figure~\ref{fig: comparison between hit percentage}) the average cost increases as we increase $(c_f)$ increases. 
\section{Conclusion}
We have formulated the optimal content fetching and caching problem~\eqref{eq:optimal caching problem} subject to hard cache capacity constraints. In Theorem~\ref{main_theorem_for_different_C_h}, we have provided the optimal policy for the single content problem with holding cost. In Theorem~\ref{thm:indexability}, we have proved the indexability of the single content problem. Finally, in Theorem~\ref{thm:whittle indices}, we have computed the Whittle indices. We have proposed a Whittle index based algorithm to solve the optimal content fetching and caching problem subject to hard cache capacity constraints. We have demonstrated that our proposed algorithm outperforms the solution provided by~\cite{9488731} and offers almost the same performance as the optimal policy. It would be interesting to design and solve the problem where content fetching can be unsuccessful; for example, if the local cache is connected to the central server or back-end cache via a wireless channel, the success of fetching content will depend upon the channel's reliability.

\bibliographystyle{ieeetr}
\bibliography{reference}
\subsection{Proof of Lemma~1} \label{appendix:proof_theorem_1}
To proof part~$(a)$, we make use of the relative value iteration algorithm~\cite[Chapter 7, Section 7.4]{bertsekas2012dynamic} as follows:
\begin{align}
   & V_{i+1}(\tau,1,1) {=}\min \left\{
     c_a\lambda \tau {+}\frac{C_h}{\beta}+\Lvtau ,  c_f{+}\frac{C_h}{\beta}+\Lvo ,\right.\nonumber \\  
     &\left. c_a\lambda \tau+V_i(0,0), c_f+V_i(0,0)   \right\} 
     {-} \min \left\{
     c_a\lambda \tau_c {+}\frac{C_h}{\beta}{+}\Lvtauc , \right. \nonumber\\
     &\left. c_f+\frac{C_h}{\beta}+\Lvo,c_a\lambda \tau_c+V_i(0,0) , c_f+V_i(0,0)   \right\}\label{v_i(tau,1,1)}\\
  & V_{i+1}(\tau,1,0) =\min \left\{\frac{C_h}{\beta}+\Lvtau,V_i(0,0)\right\} \nonumber\\
  &-\min \left\{\frac{C_h}{\beta}+\Lvtauc,V_i(0,0)\right\} \label{v_i(tau,1,0)}
\end{align}
where $V_i(s)$ is the relative cost function at $i^{th}$ iteration for state $s$, $\Lvtau:=\int_{0}^{\infty} \beta e^{-\beta t}(p V_i(t+\tau,1,1)  +(1-p)V_i(t+\tau,1,0))dt$ and $\tau_c\geq 0$ is a fixed value.  Since the embedded discrete time Markov chain has a single recurrent class, the relative value iteration algorithm~\eqref{v_i(tau,1,1)} and ~\eqref{v_i(tau,1,0)} converge to the Bellman's equations~\eqref{equation_h(tau11)} and~\eqref{equation_h(tau10)}, respectively~\cite{puterman2014markov,bertsekas2012dynamic}. Hence, it is sufficient to show that for any pair of $(\tau_1,\tau_2)$ such that $\tau_1\leq \tau_2$ and for $b\in \{0,1\}$,
\begin{align}
      V_i(\tau_1,1,b)\leq V_i(\tau_2,1,b) \forall\,\, i\in \mathbb{Z}_{+} \,\, \label{ineq: value_iteration}
\end{align}
To prove this, we follow induction method. We can define: $V_0(\tau,1,b)=0\,\forall \,\tau \geq 0$. Hence, ~\eqref{ineq: value_iteration} is satisfied for $i=0$. We assume,  $V_n(\tau_1,1,b)\leq V_n(\tau_2,1,b)$ and we prove that,  $V_{n+1}(\tau_1,1,b)\leq V_{n+1}(\tau_2,1,b)$.  \\ We note that the second term in~\eqref{v_i(tau,1,1)} and~\eqref{v_i(tau,1,0)} are independent of $\tau$.
 Since $V_n(t+\tau_1,1,0)\leq V_n(\tau_2,1,0)$ and $V_n(t+\tau_1,1,1)\leq V_n(\tau_2,1,1)$,  integrand of $L_n^{v}(\tau_1)$ is less than or equal to the integrand of $L_n^{v}(\tau_2)$. Which further implies that $L_n^{v}(\tau_1)\leq L_n^{v}(\tau_2)$. Since the second argument of the first minimization function ~\eqref{v_i(tau,1,0)} is independent of $\tau$, $V_{n+1}(\tau_1,1,0)\leq V_{n+1}(\tau_2,1,0)$ for $\tau_1\leq \tau_2.$
In a similar manner, it can be shown that $V_{n+1}(\tau_1,1,1)\leq V_{n+1}(\tau_2,1,1)$ for $\tau_1\leq\tau_2$. 
\subsection{Proof of Theorem~\ref{main_theorem_for_different_C_h}}
\label{appendix:proof_thm_3}
   To prove this theorem, we follow the following steps:
   \begin{enumerate} [(a)]
       \item We consider three mutually exclusive and exhaustive cases conditioned on the values of $h(\tau,1,1)$ and $h(\tau,1,0)$.
       \item For each cases we derive the optimal actions for those cases.
       \item Each case maps to a range of values for $C_h$. We argue that ranges of the all the cases are disjoint.
   \end{enumerate}   
 We discuss the following cases:
    \begin{enumerate}[(1)]
        \item Suppose $\frac{C_h-\theta}{\beta}+\Lot \geq h(0,0)$. \\
        In this case, $h(\tau,1,0)=h(0,0)$ and $h(0,1)=c_f+h(0,0)$ from~\eqref{equation_h(tau10)} and~\eqref{equation_h(0,1)}. To derive the value of $\theta$
we replace the value of $h(0,1)$ in~\eqref{relation_h(0,0)_h(0,1)} and hence, $h(0,0)=c_f+h(0,0)-\frac{\theta}{p \beta} $ and that further implies that $ \theta=p \beta c_f$. Since $h(\tau,1,0)$ and $h(\tau,1,1)$ are non-decreasing in $\tau$, from~\eqref{equation_h(tau11)},
\[h(\tau,1,1)=\min\{c_f+h(0,0),c_a\lambda \tau+h(0,0) \}\]
Let us define, $\tau^0=\frac{c_f}{c_a \lambda}$ and hence, $c_f<c_a\lambda\tau^{0}$ for $\tau>\tau^{0}$. Hence, $h(\tau,1,1)=\begin{cases}
    c_a\lambda \tau+h(0,0) \,\, \forall \tau \leq \tau^0 \\
    c_f+h(0,0) , \,\, \forall \tau > \tau^0 \\
\end{cases}$ \\
We summarize the optimal actions in  Table~\ref{table:case1optimalaction}.  
         \begin{table}[h]
         \centering
\caption{States and Optimal actions for case $(1)$}
\label{table:case1optimalaction}
\begin{tabular}{|c|c|c|}
\hline
    \multicolumn{3}{|c|}{Optimal actions } \\
  \hline
   $s=(\tau,1,1)$ &  $s=(\tau,1,0)$ & $s=(0,1)$\\ 
     \hline
       $\pi^{\ast}(s)=\begin{cases}
          3\text{ for } \tau \leq \tau^0 \\
          2\text{ for } \tau > \tau^0  
      \end{cases}$ & $\pi^{\ast}(s)=$ $2$ & $2$\\
     \hline
\end{tabular}
\end{table}\\
We further investigate the range of $C_h$ to satisfy the condition for $\frac{C_h-\theta}{\beta}+\Lot\geq h(0,0)$. We replace the value of $\Lot$  using $h(\tau,1,1)$ and $h(\tau,1,0)$ obtained above and get, $\frac{C_h-\theta}{\beta}+\int_{0}^{\tau^0} \beta e^{-\beta t}p c_a\lambda t+\int_{\tau^0}^{\infty} p \beta e^{-\beta t}c_f+h(0,0) \geq  h(0,0)$. After integrating and replacing the value of $\tau^{0}=\frac{c_f}{c_a\lambda}$, we obtain $\frac{C_h-p\beta c_f}{\beta}+\frac{p c_a \lambda}{\beta}(1-e^{-\beta \tau^0}) \geq  0$.  
 \begin{align}
 \text{Hence, } C_h\geq p \beta c_f-p c_a \lambda (1-e^{-\beta \tau^0}) \label{eq:cond on ch first case}
 \end{align}   
        \item Suppose,  there exists a $\bar{\tau}>0$ such that $\taub{=} \min \left\{\tau>0:\frac{C_h-\theta}{\beta}+\Lotau
    \geq h(0,0)\right \}$ and there exists a $\tau^\ast \leq \bar{\tau}$ such that
    $\taustar= \min \left\{\tau \geq 0 :c_a\lambda \tau + \Lotau = c_f+\Lot\right\} $. \\
    Since $\frac{C_h-\theta}{\beta}+\Lot < h(0,0)$, the first term is less than or equal to the second term in the minimization function~\eqref{equation_h(0,1)}. Hence, we get
    \begin{align}
    h(0,1) &=  c_f+\frac{C_h-\theta}{\beta}+\Lot  \label{h(0,1) with taubar_tautilde_case_2} 
    \end{align}
    Since $\frac{C_h-\theta}{\beta}+\Lotau\leq h(0,0)$ for $\tau \leq \taub$ the first term is less than or equal to second term in the minimization function~\eqref{equation_h(tau10)} for $\tau\leq\taub$. Hence, we get, 
    \begin{align}
        h(\tau,1,0)&=\begin{cases}
          \frac{C_h-\theta}{\beta}+\Lotau \,\, \forall \,\tau \leq \taub \\
          h(0,0)\,\, \forall \, \tau > \taub 
          \label{h_tau_0_1with taubar taustar}
        \end{cases} 
        \end{align}
         We note that the second term and fourth term in the minimization function of~\eqref{equation_h(tau11)} are constants and from the definition of $\taub$, the second term is less than or equal to the fourth term, i.e., $ c_f+\frac{C_h-\theta}{\beta}+\Lot<c_f+h(0,0)$.
        Since $\frac{C_h-\theta}{\beta}+\Lotau\leq h(0,0)$ for $\tau \leq \taub$, the first term is less than or equal to the third term in the minimization function~\eqref{equation_h(tau11)} , i.e.,  $c_a\lambda \tau +\frac{C_h-\theta}{\beta}+\Lotau \leq c_a\lambda \tau+ h(0,0)$. Hence, $h(\tau,1,1)=\min\{c_a\lambda \tau +\frac{C_h-\theta}{\beta}+\Lotau, c_f+\frac{C_h-\theta}{\beta}+\Lot\}$ for $\tau\leq\taub$. From the definition of $\taustar$, for $\tau\leq\taustar$, $c_a\lambda\tau+\Lotau \leq c_f+\Lot$. Hence, $c_a\lambda \tau +\frac{C_h-\theta}{\beta}+\Lotau \leq c_f+\frac{C_h-\theta}{\beta}+\Lot$ for $\tau \leq \taustar$. Hence, $h(\tau,1,1)=c_a\lambda \tau +\frac{C_h-\theta}{\beta}+\Lotau=c_a\lambda\tau+h(\tau,1,0)$ (from~\eqref{h_tau_0_1with taubar taustar}) for $\tau \leq \taustar$. We further note that from~\eqref{h(0,1) with taubar_tautilde_case_2}, for $\taustar<\tau\leq\taub$, $h(\tau,1,1)= c_f+\frac{C_h-\theta}{\beta}+\Lot=h(0,1)$ and since this is a constant and $h(\tau,1,1)$ is non-decreasing in $\tau$, $h(\tau,1,1)=h(0,1)$ for $\tau>\taustar$.
        \begin{align}
        h(\tau,1,1)&=\begin{cases}
            c_a \lambda \tau +h(\tau,1,0) \,\, \forall \,\tau \leq \taustar\\
            h(0,1) \,\, \forall \,\tau > \taustar \label{h_tau_1_1with taubar taustar}
        \end{cases}
    \end{align} 
     We summarize the optimal actions for all states in the Table~\ref{table:case2optimalaction}.
    \begin{table}[h]
    \centering
\caption{States and Optimal actions for case $(2)$}
\label{table:case2optimalaction}
\setlength\tabcolsep{3pt}
\begin{tabular}{|c|c|c|}
\hline
    \multicolumn{3}{|c|}{ Optimal actions} \\
  \hline 
    $s=(\tau,1,1)$ &  $s=(\tau,1,0)$ & $s=(0,1)$\\ 
     \hline
      $\pi^{\ast}(s)=\begin{cases}
         0\text{ for }\tau \leq \taustar \\
          1\text{ for } \tau > \taustar  
      \end{cases}$ & $\pi^{\ast}(s)=\begin{cases}
         0\text{ for }\tau \leq \taub \\
          2\text{ for } \tau > \taub  
      \end{cases}$ &  $1$\\
     \hline
\end{tabular}
\end{table}
In the following we obtain the expressions for $h(\tau,1,1)$ and $h(\tau,1,0).$ We subsequently show that $C_h=0$ and $\taub=\taustar$. We show that $\frac{C_h-\theta}{\beta}+\Lotau=h(0,0)$ for $\tau>\taustar$. The last observation combining with Table~\ref{table:case2optimalaction} imply that we can set $\pi^{\ast}(\tau,1,0)=0$ for all $\tau\geq 0$. We finally derive the expressions for $\taustar$ and $\theta$. \\
In the following Lemma we provide  the expressions for $h(\tau,1,1)$ and $h(\tau,1,0).$
\begin{lemma}
\label{lemma:case_2_values}
       \begin{align}&h(\tau,1,1)=\nonumber\\
        &\begin{cases}
            c_a\lambda \tau-p \beta c_a\lambda\frac{\tau^2}{2} +\theta \tau-C_h \tau+h(0,1)-c_f \text{ for $\tau \leq \taustar$}\\
            h(0,1) \text{ for $\tau > \taustar$}
        \end{cases}\label{eq:final_h(tau,1,1)} \end{align} 
          \begin{align}&\text{and }h(\tau,1,0)=\nonumber\\
        &\begin{cases}-p \beta c_a\lambda\frac{\tau^2}{2} +\theta \tau-C_h \tau+h(0,1)-c_f \text{ for $\tau \leq \taustar$}\\\frac{C_h}{p \beta}(1-e^{p \beta (\tau-\taub)})+h(0,0)\,\, \text{ for } \taustar < \tau \leq \taub \\
        h(0,0) \text{ for } \tau>\taub
        \end{cases}\label{eq:final_h(tau,1,0)}\end{align}
\end{lemma}
\begin{IEEEproof}
    See Appendix~\ref{appendix:proof_of_lemma_6}. 
\end{IEEEproof}

We observe  from~\eqref{h_tau_0_1with taubar taustar}  that $h(\taub,1,0)=h(0,0)$. This implies $ \frac{C_h-\theta}{\beta}+L_0(\taub)= h(0,0)$. We substitute the values of $h(\tau,1,1)=h(0,1)$ and $h(\tau,1,0)=h(0,0)$ for $\tau \geq \taub$ from~\eqref{h_tau_1_1with taubar taustar}
 and~\eqref{h_tau_0_1with taubar taustar}, respectively and obtain 
 $\frac{C_h-\theta}{\beta}+p h(0,1)+(1-p)h(0,0) = h(0,0)$. After replacing the value of $h(0,0)=-\frac{\theta}{p\beta}+h(0,1)$ from \eqref{relation_h(0,0)_h(0,1)}, we obtain \begin{equation} 
 C_h = 0. \end{equation} 
     We replace the value of $C_h$ in~\eqref{eq:final_h(tau,1,0)} and get, $h(\tau,1,0)=h(0,0)$ for $\taustar \leq \tau \leq  \taub$. Since $\taustar\leq \taub$ and by the definition of $\taub=\min\left\{\tau>0:\frac{C_h-\theta}{\beta}+\Lotau
    \geq h(0,0)\right \}$, $\taustar=\taub$. \\
      Recall the relative value function for the state $(\tau,1,0)$ from~\eqref{equation_h(tau10)}
    $h(\tau,1,0)=\min\left\{
          \frac{C_h-\theta}{\beta}+\Lotau,
          h(0,0)\right\}$.
       We note that for $\tau\geq \taustar,$ $\frac{C_h-\theta}{\beta}+\Lotau\stackrel{(a)}=-\frac{\theta}{\beta}+ph(0,1)+(1-p)h(0,0)=h(0,0).$ We obtain the equality (a) by replacing  $h(\tau,1,0)=h(0,0)$ and $h(\tau,1,1)=h(0,1)$ for $\tau\geq\taustar$ followed by integration. The last observation implies that for the states $s\in \{(\tau,1,0): \tau \geq \taustar\}$ the relative value function $h(\tau,1,0)$ has same value  for action $0$ and $2$, i.e., keep and discard, respectively. \\
   In the following we find the value of $\theta$ and $\taustar$ using the fact $h(\tau,1,1)=h(0,1)=c_a\lambda \taustar +h(\taustar,1,0)$  from~\eqref{h_tau_1_1with taubar taustar}. Since $\taustar=\taub$ and  $h(\taustar,1,0)=h(0,0)$ from~\eqref{h_tau_0_1with taubar taustar}, we have $ h(0,1)= c_a\lambda \taustar+h(0,0)$. By replacing the value of $h(0,0)$ from~\eqref{relation_h(0,0)_h(0,1)} we obtain $\theta = p \beta c_a\lambda \taustar$. This provides a relation between $\theta$ and $\taustar$. In the following we replace the value of $\theta$ in~\eqref{eq:final_h(tau,1,1)}. For $\tau=\taustar$, $h(\taustar,1,1)=h(0,1)=c_a\lambda\taustar+p\beta c_a\lambda\frac{{\taustar}^2}{2}+h(0,1)-c_f$. 
     Hence, we have the following quadratic equation for $\taustar$ :
    \begin{align}
        2c_a\lambda\taustar+p\beta c_a\lambda{\taustar}^2-2c_f=0\nonumber\\
        \implies {\taustar}^2+\frac{2}{p\beta}\taustar-\frac{2c_f}{c_a \lambda p\beta}=0 
    \end{align}
    Hence, $\taustar=-\frac{1}{p\beta }+\sqrt{(\frac{1}{p\beta })^2+\frac{2c_f}{p \beta c_a \lambda}}$. 
    \item  Suppose there exists a $\taub > 0$ such that $\taub{=} \min \left\{\tau>0:\frac{C_h-\theta}{\beta}+\Lotau
    \geq h(0,0)\right \} $ and there exists  a $\tilde{\tau} > \bar{\tau}$ such that $\tautil{=}\min\left\{\tau > \taub:c_a\lambda \tau + h(0,0) = c_f+\frac{C_h-\theta}{\beta}+\Lot \right\}$. We note that, the values of $h(0,1)$ and $h(\tau,1,0)$ will be same as in~\eqref{h(0,1) with taubar_tautilde_case_2} and~\eqref{h_tau_0_1with taubar taustar}, respectively. Hence, 
    \begin{align}
    h(0,1) &=  c_f+\frac{C_h-\theta}{\beta}+\Lot  \label{h(0,1) with taubar_tautilde_case_3}\\
      h(\tau,1,0)&=\begin{cases}
          \frac{C_h-\theta}{\beta}+\Lotau \,\, \forall \,\tau \leq \taub \\
          h(0,0)\,\, \forall \, \tau > \taub 
          \label{h(tau,1,0) with taubar_tautilde_case_3}
          \end{cases} \end{align}
          We note that similar to the case~$(2)$, the second term is less than or equal to the fourth term in the minimization function of~\eqref{equation_h(tau11)}, i.e., $ c_f+\frac{C_h-\theta}{\beta}+\Lot<c_f+h(0,0)$.
          To obtain the value of $h(\tau,1,1)$, we first consider the region $\tau\leq\taub$. In this case, in the minimization function of~\eqref{equation_h(tau11)}, third term is less than or equal to second term since $\tautil>\taub$ and first term is less than or equal to third term by the definition of $\taub$. Then we consider the region $\taub<\tau\leq\tautil$ and observe that in the minimization function of ~\eqref{equation_h(tau11)} the third term is less than or equal to the first term for $\tau>\taub$ by the definition of $\taub$ and third term is also less than or equal to the second term since $\tau\leq \tautil$. For $\tau>\tautil$, third term is less than or equal to first term and second term is less than of equal to third term by the definition of $\tautil$. After combining the above facts  with~\eqref{h(0,1) with taubar_tautilde_case_3} and~\eqref{h(tau,1,0) with taubar_tautilde_case_3}, 
          \begin{align}
        \text{we get, }h(\tau,1,1)&=\begin{cases}
            c_a \lambda \tau +h(\tau,1,0) \,\, \tau \leq \bar{\tau}\\
            c_a \lambda \tau + h(0,0) \,\, \bar{\tau} \leq \tau \leq \tilde{\tau} \\
            h(0,1) \,\, \tau > \tilde{\tau} \label{h(tau,1,1)with taubar_tautilde_case_3}
        \end{cases}
    \end{align}
     We summarize the optimal actions for all states in the Table~\ref{table:case3optimalaction}. 
     \begin{table}[h]
     \centering
\caption{States and Optimal actions for case $(3)$}
\setlength\tabcolsep{2pt}
\label{table:case3optimalaction}
\begin{tabular}{|c|c|c|c|}
\hline
    \multicolumn{3}{|c|}{Optimal actions} \\
  \hline 
   $s=(\tau,1,1)$ &  $s=(\tau,1,0)$ & $s=(0,1)$\\ 
     \hline
      $\pi^{\ast}(s)=\begin{cases}
          0\text{ for } \tau \leq \taub \\
           2\text{ for } \taub \leq \tau \leq \tautil \\
          1\text{ for } \tau > \tautil 
      \end{cases}$ & $\pi^{\ast}(s)=\begin{cases}
          0\text{  for } \tau \leq \taub \\
          2\text{ for }  \tau >\taub 
      \end{cases}$  &  $1$ \\
     \hline
\end{tabular}
\end{table}\\
    We further note that $\tautil \leq {\tau}^0$. Recall $\tau^{0}=\frac{c_f}{c_a\lambda}$ from case~$(1)$. $c_a\lambda\tau^{0}+h(0,0)=c_f+h(0,0)\geq c_f+\frac{C_h-\theta}{\beta}+\Lot$. Hence, by the definition of $\tautil$, $\tautil \leq \tau^{0}$. \\
    In the following we obtain the expressions for $\theta$, $h(\tau,1,1)$ and $h(\tau,1,0).$  We subsequently  derive the expressions for $\taub$ and $\tautil$. We finally derive the value for $C_h$. 
In the following Lemma we provide  the expressions for $h(\tau,1,1)$ and $h(\tau,1,0).$
\begin{lemma}
\label{lemma:case_3_values}
    \begin{enumerate}
    \item $\theta=c_a \tilde{\tau} \lambda p \beta$
        \item $h(\tau,1,0)=$ \begin{align}\begin{cases}c_a \lambda p \beta\left(\tilde{\tau} \tau-\frac{\tau^2}{2}\right)-C_h \tau+h(0,1)-c_f  \text{ for $\tau \leq \taub$}\\
        h(0,0) \text{ for } \tau>\taub
        \end{cases}\label{eq:final_h(tau,1,0)_case_3}\end{align}
        \item $h(\tau,1,1)=$\begin{align}\begin{cases}
            c_a\lambda\tau{+}c_a \lambda p \beta\left(\tilde{\tau} \tau{-}\frac{\tau^2}{2}\right){-}C_h\tau{+} h(0,1){-}c_f \text{ for } \tau {\leq} {\taub} \\
            c_a\lambda\tau+h(0,0) \text{ for } \taub<\tau\leq\tautil\\
            h(0,1) \,\,\text{for } \tau > \tilde{\tau} 
        \end{cases}\label{eq:final_h(tau,1,1)_case_3} \end{align}
    \end{enumerate}
\end{lemma}
\begin{IEEEproof}
      We derive the value of $\theta$ from~\eqref{h(tau,1,1)with taubar_tautilde_case_3} and~\eqref{relation_h(0,0)_h(0,1)}. At $\tau=\tilde{\tau},$ we have,  $c_a\lambda\tilde{\tau}+h(0,0)=h(0,1)=h(0,0)+\frac{\theta}{p \beta}$. Hence, we have $\theta=c_a \tilde{\tau} \lambda p \beta$. For the rest of the proof see Appendix~\ref{appendix:proof_of_lemma_7}
    \end{IEEEproof}
    In the following we obtain soltutions for $\taub$ and $\tautil.$
         We further note that from~\eqref{h(tau,1,0) with taubar_tautilde_case_3} and~\eqref{relation_h(0,0)_h(0,1)},
         \begin{align*}
    h(\bar{\tau},1,0)=h(0,0)&=h(0,1)-\frac{\theta}{p\beta} \end{align*}
    After replacing the value of $h(\tau,1,0)$ from Lemma~\ref{lemma:case_3_values}, we get
    \begin{align}
      c_a \lambda p \beta\left(\tilde{\tau} \taub-\frac{\taub^2}{2}\right){-}C_h \tau+h(0,1)-c_f &=h(0,1){-}\frac{\theta}{p\beta}\nonumber\\
    \implies c_a \lambda p \beta (\tilde{\tau} \taub-\frac{\taub^2}{2}){-}C_h \taub+c_a \lambda\tilde{\tau}-c_f&=0 \label{quadratic_eq_h(tau,1,0)}
         \end{align}
         We obtain the last equality after replacing the value of $\theta=c_a \tilde{\tau} \lambda p \beta$ from Lemma~\ref{lemma:case_3_values}. 
         We observe that, \eqref{quadratic_eq_h(tau,1,0)}  provides a relationship between $\tautil$ and $\taub$. However, to get closed form solutions for $\tautil$ and $\taub$, we need another equation such that solutions to these two equations provide values of $\tautil$ and $\taub$.
       We observe from~\eqref{h(tau,1,0) with taubar_tautilde_case_3}, $h(0,0)=h(\taub,1,0)= \frac{C_h-\theta}{\beta}+L_{0}(\taub)=\frac{C_h-\theta}{\beta}+e^{\beta \taub}L_{\taub}(0)$. The last equality is by change of variables in the integration. We replace the value of $h(\tau,1,0)$ for $\tau\geq\taub$ and $h(\tau,1,1)$ from~\eqref{eq:final_h(tau,1,0)_case_3} and~\eqref{eq:final_h(tau,1,1)_case_3} and we obtain $h(0,0)=$ $\frac{C_h-\theta}{\beta}{+}e^{\beta \taub}\int_{\taub}^{\tautil}\beta e^{-\beta t}p(c_a\lambda t{+}h(0,0))dt+ e^{\beta \taub}\int_{\tautil}^{\infty}\beta e^{-\beta t}p h(0,1)dt{+}(1{-}p)h(0,0)$ $=\frac{C_h-\theta}{\beta} + \frac{p c_a\lambda}{\beta}\left((1+\beta \taub)-e^{\beta (\taub-\tautil)}(1+\beta \taub)\right) + p h(0,0)(1-e^{\beta(\taub- \tautil)})+e^{\beta (\taub-\tautil)} p(\frac{\theta}{p \beta}+h(0,0))+(1-p)h(0,0)$.\\
         We replace $h(0,1)=\frac{\theta}{p\beta}+h(0,0)$ from~\eqref{relation_h(0,0)_h(0,1)}and $\theta=c_a \tilde{\tau} \lambda p \beta$ from Lemma~\ref{lemma:case_3_values} in the above equation and obtain: 
         \begin{align}
            h(0,0){=}&\frac{C_h-\theta}{\beta}+\frac{p c_a\lambda}{\beta}(1+\beta \taub)-\frac{p c_a\lambda}{\beta}e^{\beta (\taub-\tautil)}{+}h(0,0)\nonumber\\
            \implies& p c_a \lambda+C_h-p c_a \lambda \left(\beta(\tautil-\taub)+e^{-\beta(\tautil-\taub)}\right)=0 \label{exponential_eq_h(tau,1,0)}
        \end{align}
        In the following we derive the values of $C_h$.
        Recall from Lemma~\ref{lemma:uniqueness}, $\taub$ and $\tautil$ are two unique solutions of~\eqref{quadratic_eq_h(tau,1,0)} and~\eqref{exponential_eq_h(tau,1,0)}.\\ 
           {\it Lower bound on $C_h$}:
           We have $\beta(\tautil-\tau)+e^{-\beta(\tautil-\tau)}=1+\frac{ C_h}{ p c_a \lambda}$ from~\eqref{exponential_eq_h(tau,1,0)}. For $\taub=\tautil$, $C_h=0$. Since $\beta(\tautil-\tau)+e^{-\beta(\tautil-\tau)}$ is a strictly increasing function of $\tautil-\taub$, for $\tautil>\taub$, $C_h>0$. \\
           {\it Upper bound on $C_h$}:
           We show that $C_h\leq p \beta c_f-p c_a \lambda (1-e^{-\beta \tau^0})$. Recall that, $\tautil\leq \tau^0$, which  implies $\tautil-\taub\leq\tau^{0}$. Hence, it immediately follows that   $\beta (\tautil-\tau)+e^{-\beta (\tautil-\taub)}\leq \beta \tau^0+e^{-\beta \tau^0}$. After replacing $\beta (\tautil-\tau)+e^{-\beta (\tautil-\taub)}=1+
           \frac{C_h}{c_a\lambda}$ from~\eqref{exponential_eq_h(tau,1,0)}, we get, $1+
           \frac{C_h}{c_a\lambda}\leq \beta \tau^0+e^{-\beta \tau^0}$. After replacing the value of $\tau^{0}=\frac{c_f}{c_a \lambda}$, we obtain, $C_h\leq p \beta c_f-p c_a \lambda (1-e^{-\beta \tau^0})=I$ ( by definition of $I$ in Theorem~\ref{main_theorem_for_different_C_h}).  
\end{enumerate} 
   
\subsection{Proof of Lemma~\ref{lemma:case_2_values}}
\label{appendix:proof_of_lemma_6}
    We use change of variables to get, $\Lotau=e^{\beta \tau}L_{\tau}(0)$. The derivative of $e^{\beta \tau}L_{\tau}(0)$ w.r.t $\tau$ is as follows:
    \begin{align}
        \frac{d}{dt}\{e^{\beta \tau}L_{\tau}(0)\}{=}\beta e^{\beta \tau}L_{\tau}(0){-}\beta\left(ph(\tau,1,1){+}(1{-}p)h(\tau,1,0)\right) \label{eq:derivative}
    \end{align}
    From~\eqref{h_tau_0_1with taubar taustar} and~\eqref{h_tau_1_1with taubar taustar} we observe that for $\tau \leq \taustar$,\begin{align}
         h(\tau,1,1)= c_a\lambda\tau+\frac{C_h-\theta}{\beta}+\Lotau \label{eq:h(tau,1,1)_before_derivative}
    \end{align}
    To obtain the derivative of $h(\tau,1,1)$ we use~\eqref{eq:derivative} and we get, 
    \begin{align}
        \dot{h}(\tau,1,1){=}c_a\lambda+\beta e^{\beta \tau}L_{\tau}(0){-}\beta\left(ph(\tau,1,1){+}(1{-}p)h(\tau,1,0)\right)\label{eq:h(tau,1,1)_after_derivative}\end{align}
        In the following we obtain $h(\tau,1,1)$ for $\tau\leq \taustar$. For this we replace $e^{\beta \tau}L_{\tau}(0)$ from~\eqref{eq:h(tau,1,1)_before_derivative} and $h(\tau,1,0)=h(\tau,1,1)-c_a\lambda \tau$ from~\eqref{h_tau_1_1with taubar taustar} in the above equation and obtain, $\dot{h}(\tau,1,1)$
        \begin{align}
        =&c_a\lambda{-}\beta \left(h(\tau,1,1){-}(1-p)c_a\lambda\tau{-} h(\tau,1,1){+}c_a\lambda\tau {+}\frac{C_h{-}\theta}{\beta}\right) \nonumber\\
        =&c_a\lambda-p \beta c_a\lambda\tau +\theta-C_h \nonumber
    \end{align}
    The solution of the above differential equation is,  \begin{align}
        h(\tau,1,1)=&c_a\lambda \tau-p \beta c_a\lambda\frac{\tau^2}{2} +\theta \tau-C_h \tau +C_1 \text{ for } \tau \leq \taustar \label{eq:for_replacing_C_1}
        \end{align}
        Here, $C_1$ is an integration constant. To find the value of $C_1$, we use the value of $h(\tau,1,1)$ at $\tau=0$.
        \begin{align*}
       C_1= h(0,1,1)\stackrel{(a)}=\frac{C_h-\theta}{\beta}+\Lot\stackrel{(b)}=h(0,1)-c_f
        \end{align*} Where (a) and (b) follow from~\eqref{h_tau_0_1with taubar taustar} and~\eqref{h(0,1) with taubar_tautilde_case_2}, respectively.  
        After replace the value of $C_1$ in~\eqref{eq:for_replacing_C_1} from the above equation we get for $\tau \leq \taustar$, 
        \begin{align}
        h(\tau,1,1)=&c_a\lambda \tau-p \beta c_a\lambda\frac{\tau^2}{2} +\theta \tau-C_h \tau+h(0,1)-c_f \label{value_of_h(tau,1,1)_for_second_case}
    \end{align}
    After combining~\eqref{h_tau_1_1with taubar taustar} and~\eqref{value_of_h(tau,1,1)_for_second_case} we get $h(\tau,1,1)=\begin{cases}
            c_a\lambda \tau-p \beta c_a\lambda\frac{\tau^2}{2} +\theta \tau-C_h \tau+h(0,1)-c_f  \text{ for } \tau \leq \taustar\\
            h(0,1) \text{ for } \tau > \taustar
            \end{cases}$\\
            Recall that from~\eqref{h_tau_1_1with taubar taustar} $h(\tau,1,0)=h(\tau,1,1)-c_a\lambda \tau$ for $\tau\leq\taustar$. Replacing the value of $h(\tau,1,1)$ from~\eqref{value_of_h(tau,1,1)_for_second_case}, we get, \begin{align}
             h(\tau,1,0)=-p \beta c_a\lambda\frac{\tau^2}{2} +\theta \tau-C_h \tau+h(0,1)-c_f \text{ for $\tau \leq \taustar$}\label{h(tau,1,0)below_taustar}.   
            \end{align} Moreover, from~\eqref{h_tau_0_1with taubar taustar} we observe that $h(\tau,1,0)=h(0,0)$ for $\tau>\taub$. In the following, we obtain the expression of $h(\tau,1,0)$ for $\taustar<\tau\leq \taub$. For this we consider the following equation from~\eqref{h_tau_0_1with taubar taustar}.  
            \begin{align}
        h(\tau,1,0)=& \frac{C_h-\theta}{\beta}+\Lotau \stackrel{(a)}= \frac{C_h-\theta}{\beta}+ e^{\beta \tau}L_{\tau}(0) \label{eq:h(tau,1,0) before derivative}
        \end{align} where (a) follows from change of variables. We obtain the  derivative of $h(\tau,1,0)$ w.r.t. to $\tau$ using~\eqref{eq:derivative} as follows: 
        \begin{align}
            \dot{h}(\tau,1,0)=\beta e^{\beta \tau}L_{\tau}(0)-\beta\left(ph(\tau,1,1)+(1-p)h(\tau,1,0)\right) \label{eq:h(tau,1,0) after derivative}
        \end{align}
        By replacing the value of $h(\tau,1,0)=h(0,1)$ for $\tau>\taustar$ and $e^{\beta \tau}L_{\tau}(0)=h(\tau,1,0)-\frac{C_h-\theta}{\beta}$ from~\eqref{eq:h(tau,1,0) before derivative} we get,
         \begin{align}
            \dot{h}(\tau,1,0)=\theta -C_h-p \beta h(0,1)+p \beta h(\tau,1,0)\label{eq:final_h_tau,1,0_case_3}
        \end{align}
         By solving the above differential equation we get, \begin{align*}
        h(\tau,1,0)=C_2e^{p \beta \tau}+\frac{C_h-\theta+p \beta h(0,1)}{p \beta}, 
    \end{align*}
    where $C_2$ is an integration constant. 
    To find the value of $C_2$, we use the value of $h(\tau,1,0)$ at $\tau=\taub$ from~\eqref{h_tau_0_1with taubar taustar},
    \begin{align}
     h(\taub,1,0)=h(0,0)=&C_2e^{p \beta \taub}+\frac{C_h}{p \beta}+h(0,1)-\frac{\theta}{p \beta} \nonumber \\
        \implies C_2=&-\frac{C_h}{p \beta}e^{-p \beta \taub} \nonumber \\
        \text{Hence, }h(\tau,1,0)=& \frac{C_h}{p \beta}(1-e^{p \beta (\tau-\taub)})+h(0,0)\,\, \forall \taustar \leq \tau \leq \taub \label{final_value_of_h(tau,1,0)_in_case_2}
    \end{align}
    After combining~\eqref{h_tau_0_1with taubar taustar},~\eqref{h(tau,1,0)below_taustar} and~\eqref{final_value_of_h(tau,1,0)_in_case_2} we get,
    $h(\tau,1,0)=\begin{cases}-p \beta c_a\lambda\frac{\tau^2}{2} +\theta \tau-C_h \tau+h(0,1)-c_f \text{ for $\tau \leq \taustar$}\\\frac{C_h}{p \beta}(1-e^{p \beta (\tau-\taub)})+h(0,0)\,\, \text{ for } \taustar < \tau \leq \taub \\
        h(0,0) \text{ for } \tau>\taub
        \end{cases}$
        \subsection{Proof of Lemma~\ref{lemma:case_3_values}:}
        \label{appendix:proof_of_lemma_7}
        We first derive  the value of $h(\tau,1,0)$ for $\tau\leq \bar{\tau}$. By similar arguments while deriving~\eqref{eq:derivative},~\eqref{eq:h(tau,1,0) before derivative} and~\eqref{eq:h(tau,1,0) after derivative}  we get 
    
       $ \dot{h}(\tau,1,0){=}\beta e^{\beta \tau}L_{\tau}(0)-\beta \left(p h(\tau,1,1)+(1-p)h(\tau,1,0)\right).$ 
       We replace the value of $h(\tau,1,1)=c_a\lambda\tau+h(\tau,1,0)$ from~\eqref{h(tau,1,1)with taubar_tautilde_case_3} and $e^{\beta \tau}L_{\tau}(0)$ from~\eqref{eq:h(tau,1,0) before derivative} and obtain
    \begin{align}
       \dot{h}(\tau,1,0)=&\theta-C_h-\beta p c_a \lambda \tau \label{first_diff_eqn_of_h(tau,1,0)}
    \end{align}
     Solving the differential equation~\eqref{first_diff_eqn_of_h(tau,1,0)} we get,
    \begin{align}
        {h}(\tau,1,0)=\theta \tau-C_h \tau-\beta p c_a \lambda \frac{\tau^2}{2}+C_3 \label{equation_h_tau_1_0_without_value_of_theta}
        \end{align}
        where $C_3$ is an integration constant and to find the value of $C_3$, we use the following equality at $\tau=0$,
        \begin{align*}
   C_3=h(0,1,0)\stackrel{(a)}=\frac{C_h-\theta}{\beta}+\Lot \stackrel{(b)}=h(0,1)-c_f \end{align*}  where equalities~(a) and~(b) follow from~\eqref{h(tau,1,0) with taubar_tautilde_case_3} and ~\eqref{h(0,1) with taubar_tautilde_case_3},respectively.            After replacing the value of $C_3$ in~\eqref{equation_h_tau_1_0_without_value_of_theta} we obtain \begin{align}
       {h}(\tau,1,0)=\theta \tau-C_h \tau-\beta p c_a \lambda \frac{\tau^2}{2}+h(0,1)-c_f\label{eq:41}
   \end{align}
   Substituting the value of $\theta=c_a \tilde{\tau} \lambda p \beta$ in~\eqref{eq:41} we get,
        \begin{align}
          h(\tau,1,0)=&c_a \lambda p \beta\left(\tilde{\tau} \tau-\frac{\tau^2}{2}\right)-C_h \tau+h(0,1)-c_f \nonumber
         \end{align} 
         The above expression provides the value of $h(\tau,1,0)$ for $\tau\leq\taub$. We combine this with~\eqref{h(tau,1,0) with taubar_tautilde_case_3} and obtain 
         $h(\tau,1,0)=\begin{cases}c_a \lambda p \beta\left(\tilde{\tau} \tau-\frac{\tau^2}{2}\right)-C_h \tau+h(0,1)-c_f  \text{ for $\tau \leq \taub$}\\
        h(0,0) \text{ for } \tau>\taub
        \end{cases}$.\\ We combine the value of $h(\tau,1,0)$ with~\eqref{h(tau,1,1)with taubar_tautilde_case_3} and obtain $h(\tau,1,1)=\begin{cases}
            c_a\lambda\tau+c_a \lambda p \beta\left(\tilde{\tau} \tau-\frac{\tau^2}{2}\right)-C_h + h(0,0)-c_f \,\, \text{ for } \tau \leq {\taub} \\
            c_a\lambda\tau+h(0,0) \text{ for } \taub<\tau\leq\tautil\\
            h(0,1) \,\,\text{for } \tau > \tilde{\tau} 
        \end{cases}$. 
        \subsection{Proof of Lemma~\ref{lemma:uniqueness}:}
        \label{proof_of_uniqueness_lemma}
        \begin{enumerate}

       \item   Let $\beta(\tautil-\taub):=x$. Hence, we can rewrite~\eqref{exponential_eq_h(tau,1,0)} as $x+e^{-x}=1+\frac{C_h}{ p c_a \lambda}$.
         We note that $x+e^{-x}$ is an increasing monotone function of $x$. Hence~\eqref{exponential_eq_h(tau,1,0)} has a unique solution. Furthermore,  whenever $C_h=0$, $x=0$ and $x>0$ for $C_h>0$, and as $C_h$ increases $x$ also increases. Let the solution of $x+e^{-x}=1+\frac{C_h}{ p c_a \lambda}$ be $\fng$, where $g$ is some function. Although, it is a function of $c_a,C_h,\lambda,\text{ and }p$ . Since $c_a, \lambda$ and $p$ are fixed and $C_h$ can be varied, with abuse of notation we use $\fng$,   Hence, $\beta(\tautil-\taub)=\fng \implies \tautil=\frac{\fng}{\beta}+\taub$. Consider~\eqref{quadratic_eq_h(tau,1,0)} at $\taub$ and replace the value of $\tautil$, and we get,
        \begin{align}
                &c_a \lambda p \beta \left(\frac{\fng}{\beta}\taub{+}\frac{\taub^2}{2}\right)\nonumber{-}C_h \taub{+}c_a \lambda(\frac{\fng}{\beta}+\taub){-}c_f{=}0 \nonumber\\     &\taub^2{+}2\taub(\frac{\fng}{\beta}{-}\frac{C_h}{c_a\lambda p\beta}{+}\frac{1}{p\beta})
    {-}2\left(\frac{c_f}{c_a\lambda p \beta}{-}\frac{\fng}{p\beta^2}\right)=0\nonumber
           \end{align}
           Hence, $\taub=-\left(\frac{\fng}{\beta}-\frac{C_h}{c_a\lambda p\beta}+\frac{1}{p\beta}\right)\underset{-}{+}\sqrt{\left(\frac{\fng}{\beta}-\frac{C_h}{c_a\lambda p\beta}+\frac{1}{p\beta}\right)^2+2\left(\frac{c_f}{c_a\lambda p \beta}-\frac{\fng}{p\beta^2}\right)}$.\\
           Since $\fng+e^{-\fng}=1+ \frac{C_h}{pc_a\lambda}$ or $\fng-\frac{C_h}{pc_a\lambda}=(1-e^{-\fng})$. We make the following observations: \begin{enumerate}[(a)]
           \item Since $\fng\geq0$ we observe from the above equation that $\fng - \frac{C_h}{pc_a\lambda}\geq 0$ or $\frac{\fng}{\beta}-\frac{C_h}{c_a\lambda p\beta}\geq 0$.
           \item Since $\fng$ increases as $C_h$ increases, $\fng-\frac{C_h}{pc_a\lambda}$ increases as $C_h$ increases.
           \end{enumerate} Furthermore,  we note that  since $\tautil\leq \tau^0$, $\fng\leq \beta \tau^{0}$. Which further implies that $\frac{\fng}{\beta}\leq \frac{c_f}{c_a\lambda}$ as $\tau^{0}=\frac{c_f}{c_a\lambda}$. The last observation combining with~$(b)$ implies that $\taub$ has a unique non-negative solution as follows:
            $\taub=-\left(\frac{\fng}{\beta}-\frac{C_h}{c_a\lambda p\beta}+\frac{1}{p\beta}\right)+\sqrt{\left(\frac{\fng}{\beta}-\frac{C_h}{c_a\lambda p\beta}+\frac{1}{p\beta}\right)^2+2\left(\frac{c_f}{c_a\lambda p \beta}-\frac{\fng}{p\beta^2}\right)}$. Since $\tautil\geq\taub$, $\tautil$ is also unique non-negative value. 
            \item In the following we show that $\taub$ is decreasing function of $C_h$. Since $\fng\geq0$, we obtain an upper bound of $\taub\leq-\left(\frac{\fng}{\beta}-\frac{C_h}{c_a\lambda p\beta}+\frac{1}{p\beta}\right)+\sqrt{\left(\frac{\fng}{\beta}-\frac{C_h}{c_a\lambda p\beta}+\frac{1}{p\beta}\right)^2+\frac{2c_f}{c_a\lambda p \beta}}$. Since $\fng-\frac{C_h}{pc_a\lambda}$ increases as $C_h$ increases (from~$(b)$), the R.H.S is a decreasing function of $C_h$. Since $\fng$ is an increasing function of $C_h$, if we use $2\left(\frac{c_f}{c_a\lambda p \beta}-\frac{\fng}{p\beta^2}\right)$ instead of $\frac{2c_f}{c_a\lambda p \beta}$ inside the square root of the R.H.S, it further decreases as $C_h$ increases. We note that, the modified R.H.S is $\taub$ and is a decreasing function of $C_h$. Recall that at $C_h=0$, $\fng=0$ and this implies that $\tautil=\taub=\taustar$ at $C_h=0$. Hence, we obtain $\taub\leq\taustar$ and $\taub$ is a decreasing function of $C_h$. In the following we show that $\tautil$ is an increasing function of $C_h$ and to see this we need the following characterization of $\fng$. 
           \paragraph*{Characterization of $\fng$}
           Let $C_w=1+\frac{C_h}{pc_a\lambda}$ and recall that we defined $x=\beta(\tautil-\tau)$ Hence, we can rewrite~\eqref{exponential_eq_h(tau,1,0)} as \begin{align*}
              & x+e^{-x}=C_w \implies x-C_w =- e^{-x}\\
              &{\implies} (x-C_w)e^x=-1
               {\implies} (x-C_w)e^{(x-C_w)}=-e^{-C_w}\nonumber
           \end{align*}
           Hence, $x=LambertW(-e^{-C_w})+C_w$ or $x=W(-e^{-C_w})+C_w=\fng$ (by definition). $LambertW(-e^{-C_w})$ is differentiable w.r.t $-e^{-C_w}$ for $\{\ C_w :-e^{-C_w}\notin \{0,-\frac{1}{e}\} \}$~\cite{corless1996lambert}. This further implies that, $\fng$ is differentiable w.r.t. $C_h$ for $C_h>0$ since at $C_h=0$, $-e^{-C_w}=-\frac{1}{e}$.    
         \begin{align}
               &\text{ Recall, }\tautil=\frac{\fng}{\beta}+\taub =\left(\frac{C_h}{c_a\lambda p\beta}-\frac{1}{p\beta}\right)\nonumber\\
               &+\sqrt{\left(\frac{\fng}{\beta}-\frac{C_h}{c_a\lambda p\beta}+\frac{1}{p\beta}\right)^2+2\left(\frac{c_f}{c_a\lambda p \beta}-\frac{\fng}{p\beta^2}\right)}
           \end{align}
           We differentiate $\tautil$ w.r.t. $C_h$ for $C_h>0$ and obtain
           \begin{align}
              & \frac{d\tautil}{dC_h}=\frac{1}{c_a\lambda p\beta}+\nonumber\\
            &\frac{\left(\frac{\fng}{\beta}-\frac{C_h}{c_a\lambda p\beta}\right)\frac{\gprime}{\beta}{-}\frac{1}{c_a\lambda p\beta}\left(\frac{\fng}{\beta}-\frac{C_h}{c_a\lambda p\beta}+\frac{1}{p\beta}\right)}{\sqrt{\left(\frac{\fng}{\beta}-\frac{C_h}{c_a\lambda p\beta}+\frac{1}{p\beta}\right)^2+2\left(\frac{c_f}{c_a\lambda p \beta}-\frac{\fng}{p\beta^2}\right)}}\nonumber\\
               &\geq \frac{\left(\frac{\fng}{\beta}-\frac{C_h}{c_a\lambda p\beta}\right)\frac{\gprime}{\beta}}{\sqrt{\left(\frac{\fng}{\beta}-\frac{C_h}{c_a\lambda p\beta}+\frac{1}{p\beta}\right)^2+2\left(\frac{c_f}{c_a\lambda p \beta}-\frac{\fng}{p\beta^2}\right)}} \geq 0
           \end{align}
           where $\gprime$ is the derivative of $\fng$ w.r.t. $C_h$.  The second last inequality follows from the fact that $\left(\frac{\fng}{\beta}-\frac{C_h}{c_a\lambda p\beta}+\frac{1}{p\beta}\right)\leq \sqrt{\left(\frac{\fng}{\beta}-\frac{C_h}{c_a\lambda p\beta}+\frac{1}{p\beta}\right)^2+2\left(\frac{c_f}{c_a\lambda p \beta}-\frac{\fng}{p\beta^2}\right)}$. As mentioned earlier as $C_h$ increases $\fng$ increases, therefore $\gprime$ is positive for $C_h>0$. Thus the last inequality follows since $\fng - \frac{C_h}{pc_a\lambda}\geq 0$ from~$(a)$. Hence $\tautil$ increases as $C_h$ increases. Thus $\taub\leq\taustar\leq\tautil.$
            \end{enumerate}
\end{document}